\documentclass[11pt]{article}

\usepackage{amsmath}
\usepackage{times}
\usepackage{graphicx}
\usepackage{amssymb}
\usepackage{amsthm}
\usepackage{textcomp}
\usepackage[T1]{fontenc}

\begin{document}

\newtheorem{theorem}{Theorem}
\newtheorem{lemma}{Lemma}
\newtheorem{corollary}{Corollary}
\newtheorem{definition}{Definition}
\newtheorem{conjecture}{Conjecture}
\newtheorem{identity}{Identity}
\newtheorem{axiom}{Axiom}
\newtheorem{problem}{Problem}
\newtheorem{prop}{Proposition}
\newtheorem*{axiom0}{Axiom 0}
\newtheorem*{axiom1}{Axiom 1}

\newcommand{\Real}{{\bf R}}
\newcommand{\tr}{{\mbox{tr}}}
\newcommand{\real}{{\bf R}}

\title{On Wave Dark Matter, \\ Shells in Elliptical Galaxies, \\ and the Axioms of General
Relativity}
\author{Hubert L. Bray
\thanks{Mathematics and Physics Departments, Duke University, Box 90320, Durham, NC  27708, USA, bray@math.duke.edu}}
%\date{June 21, 2012}
\maketitle

\begin{abstract}

This paper is a sequel to the author's paper entitled ``On Dark Matter, Spiral Galaxies, and
the Axioms of General Relativity'' which explored a geometrically natural axiomatic definition for dark matter modeled by a scalar field satisfying the Einstein-Klein-Gordon wave equations which, after much calculation, was shown to be consistent with the observed spiral and barred spiral patterns in disk galaxies, as seen in Figures \ref{SG1}, \ref{SG2}, \ref{SG3}, \ref{SG4}.  We give an update on where things stand on this ``wave dark matter'' model of dark matter (aka scalar field dark matter and boson stars), an interesting alternative to the WIMP model of dark matter, and discuss how it has the potential to help explain the long-observed interleaved shell patterns, also known as ripples, in the images of elliptical galaxies.

\end{abstract}

\begin{figure}
   \begin{center}
   \includegraphics[height=72mm]{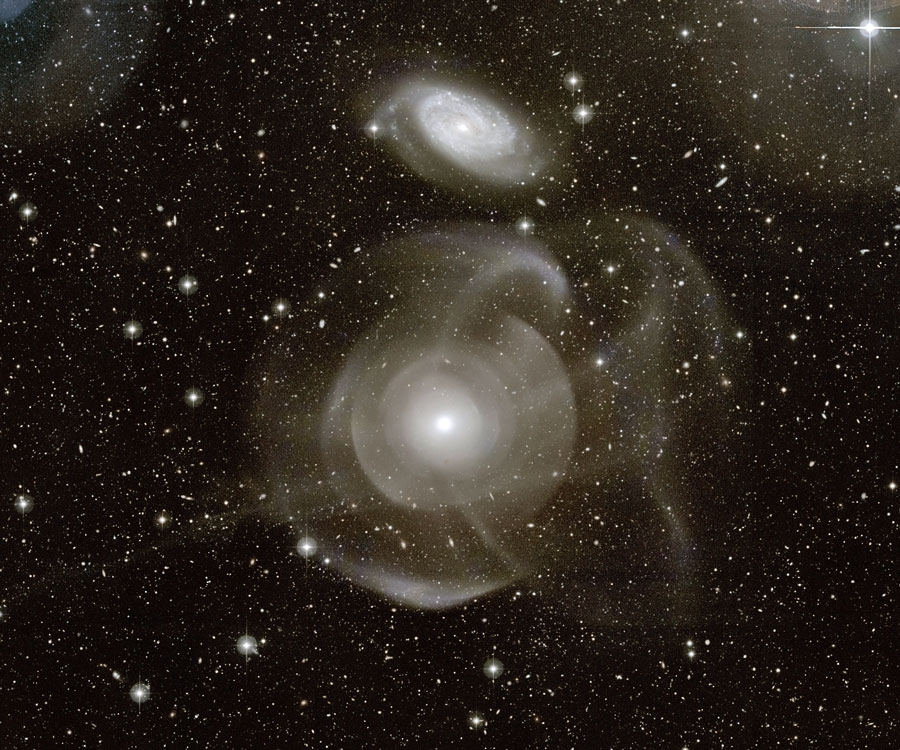}
   \includegraphics[height=72mm]{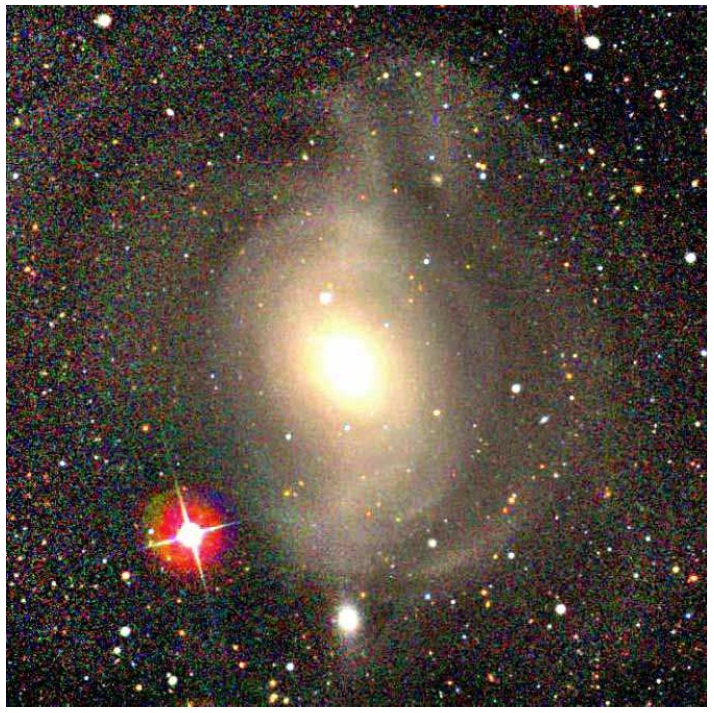}
   \end{center}
   \caption{Shells in Elliptical Galaxies:   NGC 474 (left) and NGC 4382 (right) are two examples of
   elliptical galaxies with shells that are prominent.
            Left Image Credit and Copyright: P.-A. Duc (CEA, CFHT), Atlas 3D Collaboration.
%            Right Image Credit and Copyright: ???
            } \label{F1}
\end{figure}

In section \ref{intro}, we begin with a discussion of dark matter and how the wave dark matter model  compares with observations related to dark matter, particularly on the galactic scale.  In section \ref{shells}, we show explicitly how wave dark matter shells may occur in the wave dark matter model via approximate solutions to the Einstein-Klein-Gordon equations.  How much these wave dark matter shells (as in Figure \ref{F2}) might contribute to visible shells in elliptical galaxies (as in Figure \ref{F1}) is an important open question.

\section{Introduction}\label{intro}

What is dark matter?  No one really knows exactly, in part because dark matter does not interact significantly with light, making it invisible.  Hence, everything that is known about dark matter is due to its gravity, which is significant since it composes 23\% of the mass of the universe \cite{WMAP}.  ``Regular'' matter which makes up the periodic table, typically referred to as ``baryonic matter,'' composes only about 5\% of the mass of the universe \cite{WMAP}.  Dark energy, not to be confused with dark matter, is predicted by and is one of the great successes of general relativity.  It is a very small but positive energy density with negative pressure spread evenly throughout the universe in a manner that explains the observed accelerating expansion of the universe \cite{AU1,AU2}.  Dark energy composes about 73\% of the mass of the universe \cite{WMAP}.  However, since dark energy is evenly spread throughout the universe, it is an insignificant component of the mass of galaxies.

Dark matter, on the other hand, is {\it most} of the mass of galaxies \cite{BM, BT}.  These dark matter ``halos'' in galaxies are roughly spherically symmetric even when the visible baryonic matter is mostly in a disk \cite{BM, BT}.  Hence, while spiral galaxies are disk-like in terms of the light they emit, by mass they are roughly spherically symmetric.  Elliptical galaxies are also dominated by these dark matter halos.  For a summary of what is known about dark matter, see \cite{DMAW,BHS,Hooper,Trimble,Ostriker}.

\begin{figure}
   \begin{center}
   \includegraphics[height=75mm]{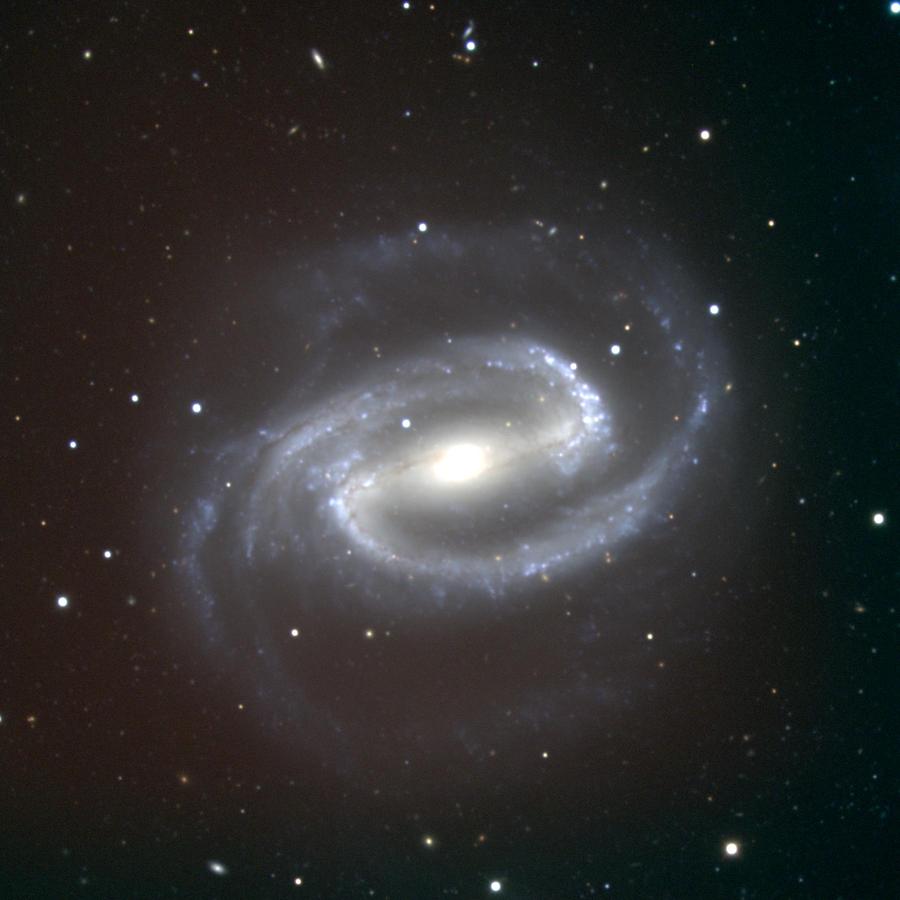}
   \includegraphics[height=75mm]{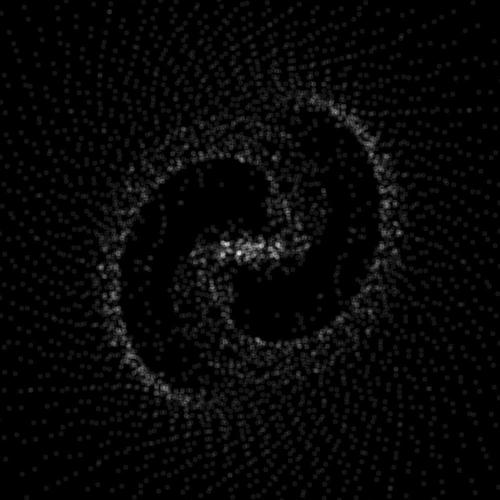}
   \end{center}
   \caption{NGC 1300 on the left, wave dark matter simulation \cite{DMSG} on the right (showing gas, dust, and stars, not dark matter).
   Left photo credit: Hillary Mathis/NOAO/AURA/NSF.  Date:  December 24, 2000. Telescope:  Kitt Peak National Observatory's 2.1-meter telescope.
   Image created from fifteen images taken in the BVR pass-bands. } \label{SG1}
\end{figure}

It may be reasonable to look at detailed images of galaxies to understand dark matter since these are dark matter dominated systems.  The two main types of galaxies are disk galaxies, most of which are spiral galaxies, and elliptical galaxies.  The common occurrence of spiral galaxies could be a major hint as to the nature of dark matter.  The author explored this idea in detail in \cite{DMSG} and produced simulated images of disk
galaxies with spiral and barred spiral patterns very similar to photos of actual galaxies, as can be seen in Figures \ref{SG1}, \ref{SG2}, \ref{SG3}, and \ref{SG4}.

\subsection{Shells in Elliptical Galaxies}

As spectacular as spiral patterns in disk galaxies are, there is also a fascinating phenomenon in the images of elliptical galaxies - the common occurrence of interleaved shells, also known as ripples, in their images.  While shells in elliptical galaxies are usually hard to see with the unaided eye, computer enhancement of photos of elliptical galaxies clearly show partial circular steps in the luminosity of the elliptical galaxies typically on the order of 3\% to 5\% of the galaxy's surface brightness profile \cite{BM}.  These shells have a global structure within the galaxy in that the centers of the partial circular shells lie at the galactic center.  In addition, the radii of the shells on the two opposite sides of the galaxies are typically interleaved, in that the radii of the shells in increasing order alternate between two sides of the galaxy as explained on page 202 of \cite{BM}.  These shells, present in 10\% to 20\% of all elliptical galaxies \cite{BM}, or perhaps 30\% to 50\% ``depending on how closely one looks'' \cite{BT}, have been known to exist since 1980.  See Figure \ref{F1} for two prominent examples of shells in elliptical galaxies.  Quoting from \cite{BM}, page 203:

\begin{figure}
   \begin{center}
   \includegraphics[height=59mm]{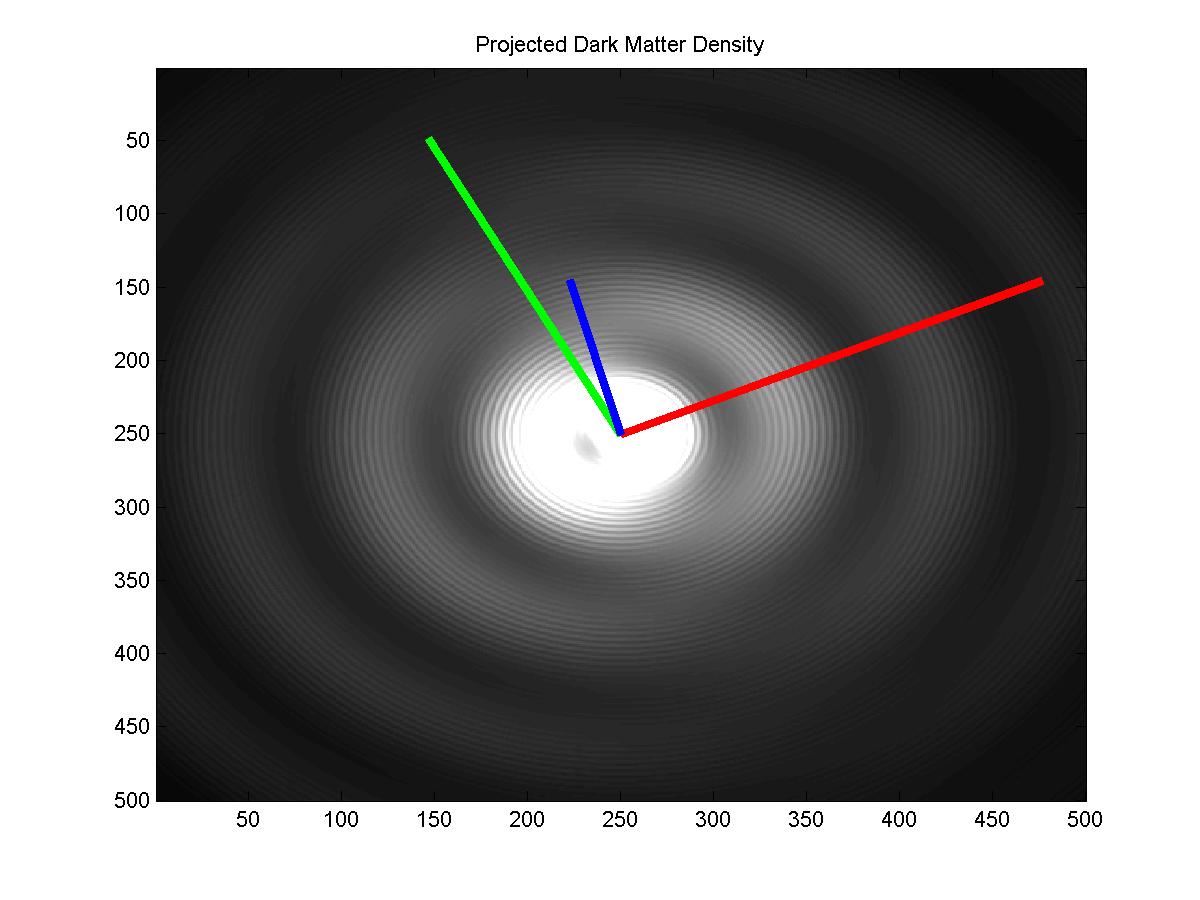}
   \includegraphics[height=59mm]{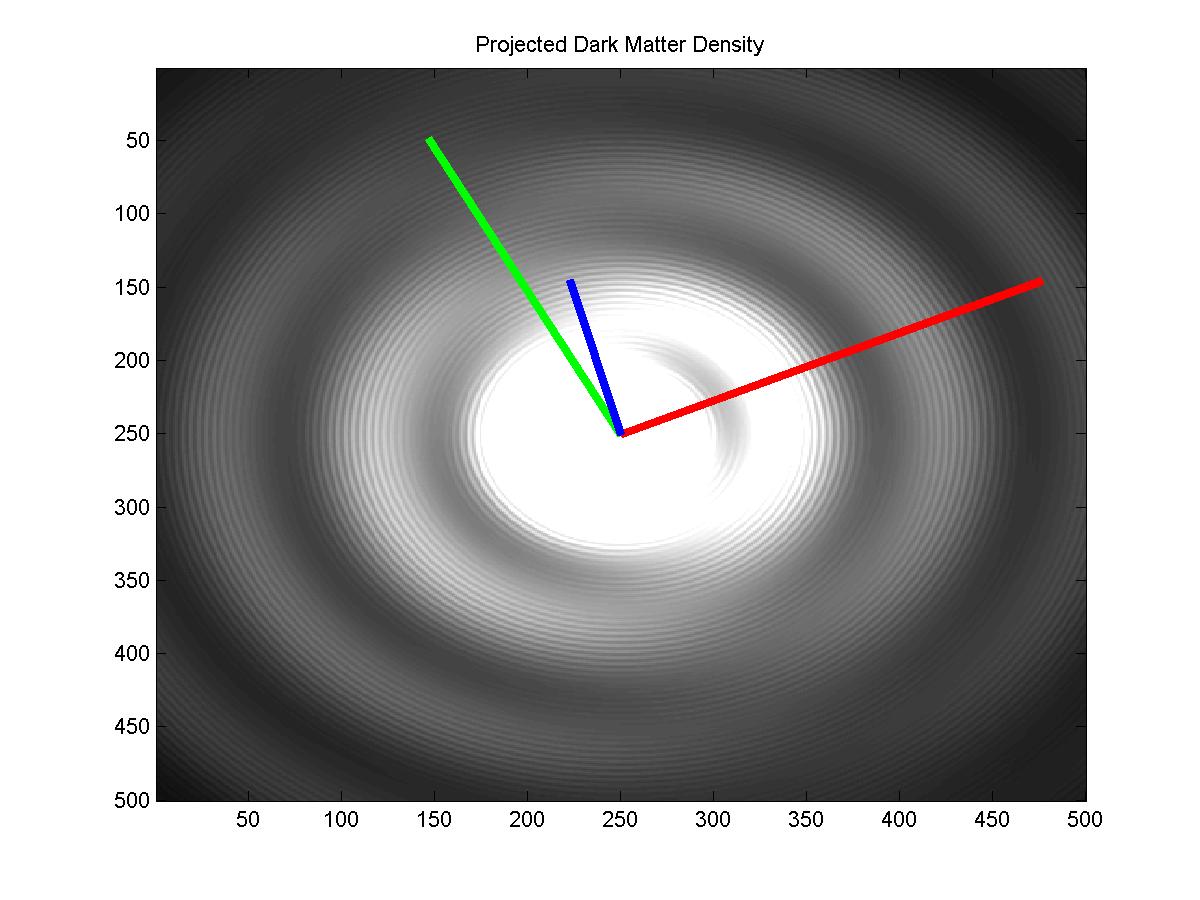}
   \end{center}
   \caption{Shells in Elliptical Galaxies:  Simulated images of wave dark matter density projected
    into the plane clearly show interleaved shells which result from first degree spherical harmonic perturbations from spherically symmetric scalar field solutions to the Klein-Gordon equation.} \label{F2}
\end{figure}

\begin{quotation}
In the classical dynamical model of an elliptical, phase space is populated very smoothly.  Therefore, the existence of ripples directly challenges the classical picture of ellipticals.  One likely possibility is that ellipticals acquire ripples late in life as a result of accreting material from a system within which there are relatively large gradients in phase-space density.  Systems with large density gradients in phase-space include disk galaxies and dwarf galaxies:  in a thin disk, the phase-space density of stars peaks strongly around the locations of circular orbits, while in a dwarf galaxy all stars move at approximately the systemic velocity, so that there is only a small spread in velocity space.

Numerical simulations suggest that ripples can indeed form when material is accreted from either a disk galaxy or a dwarf system - see Barnes and Hernquist (1992) for a review.  Moreover, simulations have successfully reproduced the interleaved property of ripples described above.  Despite these successes, significant uncertainties still surround the ripple phenomenon because the available simulations have important limitations, and it is not clear how probably their initial conditions are.
\end{quotation}

Implicit in most if not all studies of shells and ripples in elliptical galaxies to date is that dark matter is correctly modeled as a WIMP (Weakly Interacting Massive Particle).  In this paper as in the previous paper \cite{DMSG}, we study a geometrically natural axiomatic definition for dark matter, called ``wave dark matter,'' as an interesting alternative model of dark matter.  Furthermore, in section \ref{shells} of this paper we show how this geometric model for dark matter may form interleaved shells in its galactic density profile, as shown in Figure \ref{F2}.

We suggest that this global interleaved shell structure in the wave dark matter exemplified in Figure \ref{F2}, while invisible itself, could conceivably play a role in the formation of visible interleaved shells in some elliptical galaxies, perhaps as exemplified in Figure \ref{F1}, through gravity, friction, dynamical friction, and the processes mentioned in the above quote.  This conjecture is based primarily on the striking qualitative similarity between the images in Figures \ref{F1} and \ref{F2}, namely that both sets of images have interleaved shells.  The basic idea that regions of increased dark matter density could gravitationally attract more regular visible matter and hence contribute to visible shells is plausible but is also a very subtle question to study.
Resolving this conjecture with a high degree of confidence may require sophisticated computer simulations based on expert modelings of many different astrophysical processes, making this a challenging conjecture to study carefully.

The primary goal of this paper is to show how interleaved shells may occur in approximate wave dark matter solutions to the Einstein-Klein-Gordon equations, done in section \ref{shells}.  How much these wave dark matter shells (as in Figure \ref{F2}) might contribute to visible shells in elliptical galaxies (as in Figure \ref{F1}) is an important open question.

\subsection{Geometric Motivation for Wave Dark Matter}

 The only reason dark matter is known to exist is because of its gravity, which affects the orbits of stars and causes gravitational lensing of light.  But according to general relativity, gravity is just an effect of the curvature of spacetime.  Hence, at this point, dark matter could just as legitimately be called ``unexplained curvature'' on the scale of galaxies and higher.

To this end, in \cite{DMSG} we reexamined the question of what the fundamental assumptions of general relativity are exactly.  In doing so, we defined two axioms, called Axiom 0 and Axiom 1 in \cite{DMSG}.  At the highest level, one could say that

\begin{itemize}
{\item Axiom 0 = ``Geometry describes the fundamental objects of the universe.''}
{\item Axiom 1 = ``Analysis describes the fundamental laws of the universe.''}
\end{itemize}

\begin{figure}
   \begin{center}
   \includegraphics[height=75mm]{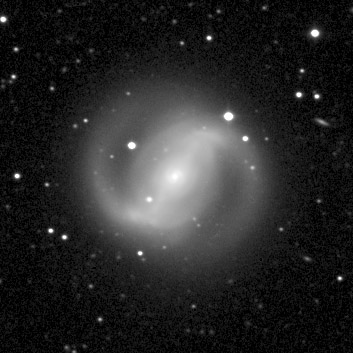}
   \includegraphics[height=75mm]{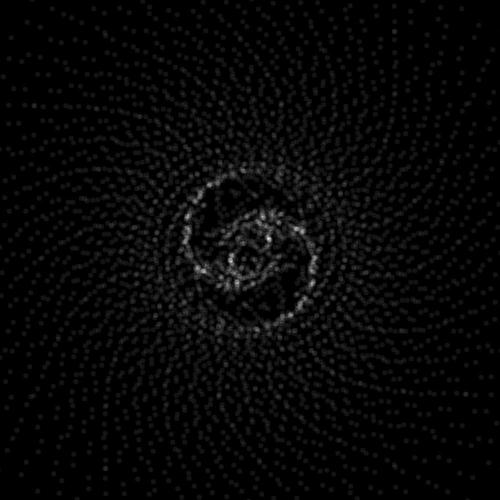}
   \end{center}
   \caption{NGC 4314 on the left, wave dark matter simulation \cite{DMSG} on the right (showing gas, dust, and stars, not dark matter).
   Left photo credit:  G. Fritz Benedict, Andrew Howell, Inger Jorgensen, David Chapell (University of Texas),
   Jeffery Kenney (Yale University), and Beverly J. Smith (CASA, University of Colorado), and NASA.  Date:  February 1996.
   Telescope:  30 inch telescope Prime Focus Camera, McDonald Observatory.}\label{SG2}
\end{figure}

Others may have their own ideas about how to take these two provocative statements and turn them into precise axioms which result in general relativity and generalizations of general relativity. The author played this game and came up with his own axioms in \cite{DMSG} which result in general relativity with a cosmological constant (dark energy) and a scalar matter field (dark matter perhaps?) described by the Klein-Gordon equation.  Regular baryonic matter is not part of the core theory resulting from the axioms but may be added in manually.

It turns out that a scalar field model of dark matter is not ruled out at this time as a viable theory of dark matter.  In fact, astrophysicists, beginning with a quantum theory motivation, have already been studying these equations as a possible model for dark matter \cite{Sin1992,JS,JWLee,LeeKoh1992,LeeKoh1996,Flat,MSBS,SM,SKM,GM}.

This ``scalar field dark matter'' model also goes by the name ``boson stars.''
In this paper, we will refer to this model of dark matter by a new name:  ``wave dark matter.''  We prefer ``wave dark matter'' because the scalar field representing the dark matter satisfies the Klein-Gordon equation, a wave equation on the spacetime.  Hence, the mental picture one should have for the qualitative characteristics for wave dark matter are waves on a pond, but with gravity and one more very important difference:  Unlike waves on a pond which have a single characteristic velocity, group velocities of solutions to the Klein-Gordon equation may be anything less than the speed of light, with smaller group velocities for long wavelengths and larger group velocities for shorter wavelengths.  In addition, these waves may interfere with one another both constructively and destructively.  Thus, if we are looking for a name for this model of dark matter which is descriptive of how it behaves, then ``wave dark matter'' is a natural choice.

\begin{figure}
   \begin{center}
   \includegraphics[height=75mm]{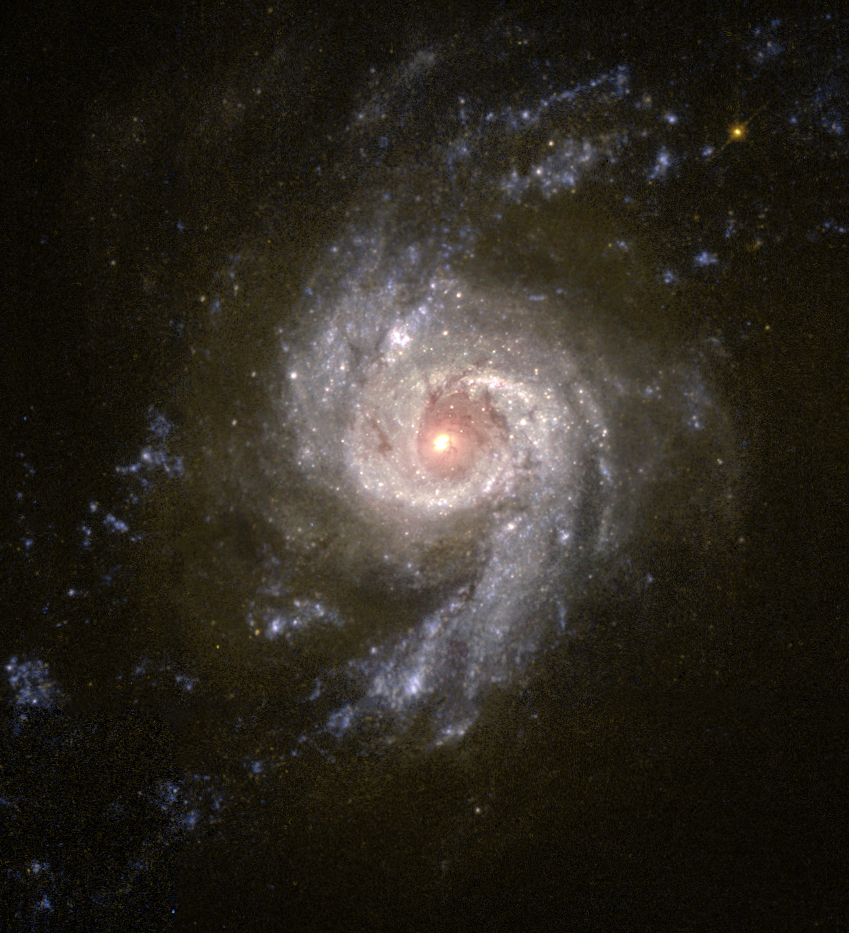}
   \includegraphics[height=75mm]{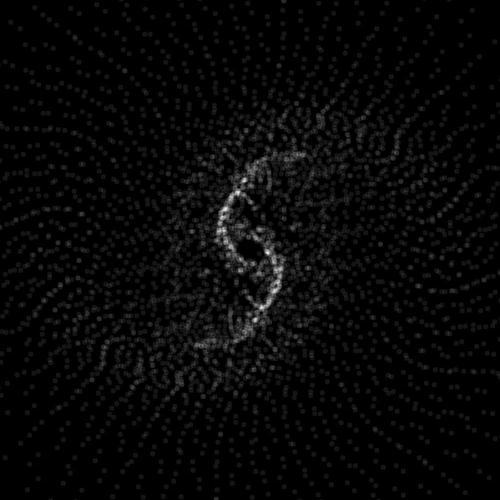}
   \end{center}
   \caption{NGC 3310 on the left, wave dark matter simulation \cite{DMSG} on the right (showing gas, dust, and stars, not dark matter).  Left photo credit:
   NASA and The Hubble Heritage Team (STScI/AURA).  Acknowledgment: G.R. Meurer and T.M. Heckman (JHU), C. Leitherer, J. Harris and
   D. Calzetti (STScI), and M. Sirianni (JHU).  Dates:  March 1997 and September 2000.  Telescope:  Hubble Wide Field Planetary Camera 2.} \label{SG3}
\end{figure}

The following axiom is one way to express the idea that ``Geometry describes the fundamental objects of the universe.''

\begin{axiom0}\label{A0}
The fundamental objects of the universe are a smooth spacetime manifold $N$, which is both Hausdorff and second countable, with a smooth metric $g$ of signature $(-+++)$ and a smooth connection $\nabla$ defined on the interior of $N$.
\end{axiom0}

A smooth manifold $N$ is a Hausdorff space with a complete atlas of
smoothly overlapping coordinate charts \cite{ONeill}.  Hence, we are reminded
that coordinate charts are more than convenient places to do
calculations, but are in fact a necessary part of the definition of
a smooth manifold. Given a fixed coordinate chart, let
$\{\partial_i\}$, $0 \le i \le 3$, be the tangent vector fields to
$N$ corresponding to the standard basis vector fields of the
coordinate chart. Let $g_{ij} = g(\partial_i, \partial_j)$ and
$\Gamma_{ijk} = g( \nabla_{\partial_i} \partial_j,
\partial_k )$, and let
\[
M = \{g_{ij}\} \;\;\;\mbox{ and }\;\;\; C = \{\Gamma_{ijk}\}
\;\;\;\mbox{ and }\;\;\; M' = \{g_{ij,k}\} \;\;\;\mbox{ and }\;\;\;
C' = \{\Gamma_{ijk,l}\}
\]
be the components of the metric and the connection in the coordinate
chart and all of the first derivatives of these components in the
coordinate chart.

The analyst in all of us knows that given p.d.e.s on a manifold, even tensorial p.d.e.s which do not depend on choices of coordinate charts, the existence and regularity theory for those p.d.e.s, which is crucial, reduces to understanding them expressed in some choice of coordinate chart.  Hence, in the spirit of searching for the easiest theories first, one way to guarantee second order p.d.e.s in coordinate charts is to require the action to be no worse than first order quadratic in coordinate charts.  These analytical considerations inspired the following interpretation of ``Analysis describes the fundamental laws of the universe.''

\begin{figure}
   \begin{center}
   \includegraphics[height=75mm]{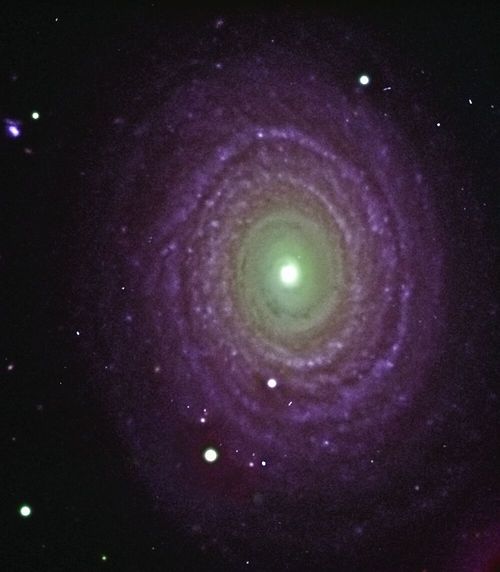}
   \includegraphics[height=75mm]{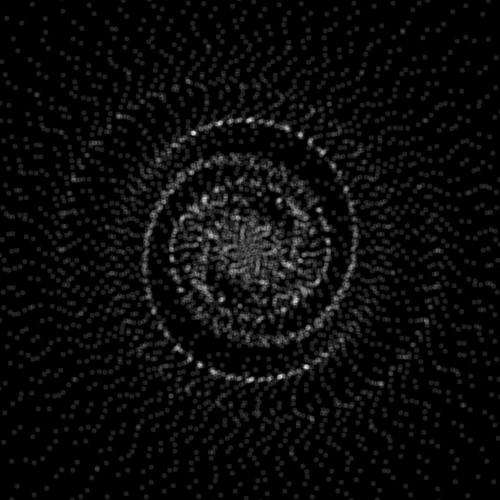}
   \end{center}
   \caption{NGC 488 on the left, wave dark matter simulation \cite{DMSG} on the right (showing gas, dust, and stars, not dark matter).  Left photo credit:
   Johan Knapen and Nik Szymanek.  Telescope:  Jacobus Kapteyn Telescope. B, I, and H-alpha bands.}\label{SG4}
\end{figure}

\begin{axiom1}\label{A1}
For all coordinate charts $\Phi : \Omega \subset N \rightarrow R^4$
and open sets $U$ whose closure is compact and in the interior of
$\Omega$, $(g,\nabla)$ is a critical point of the functional
\begin{equation}
   F_{\Phi,U}(g,\nabla) = \int_{\Phi(U)}
   \mbox{Quad}_M(M' \cup M \cup C' \cup C)
   \; dV_{R^4}
\end{equation}
with respect to smooth variations of the metric and connection
compactly supported in $U$, for some fixed quadratic functional
$Quad_M$ with coefficients in $M$.
\end{axiom1}
Note that we have not specified the action, only the form of the
action.  As is standard, we define
\begin{equation}
   \mbox{Quad}_{Y}(\{x_\alpha\}) = \sum_{\alpha,\beta}
   F^{\alpha\beta}(Y) x_\alpha x_\beta
\end{equation}
for some functions $\{F^{\alpha\beta}\}$ to be a quadratic
expression of the $\{x_\alpha\}$ with coefficients in $Y$.

The implications of the above axioms when the connection is assumed to be the Levi-Civita connection
have been long understood.  When the integrand in Axiom 1 is
reduced to $Quad_M(M')$, vacuum general relativitiy generically
results. When the integrand is reduced to $Quad_M(M' \cup M)$,
vacuum general relativity with a cosmological constant generically
results. Here ``generically'' means for a generic choice of
quadratic functional, so that the zero quadratic function, for
example, is not included in these claims. These two results were
effectively proved by Cartan \cite{Cartan}, Weyl \cite{Weyl}, and
Vermeil \cite{Vermeil} and pursued further by Lovelock \cite{Lovelock}.
The point to keep in mind is that since $(g,\nabla)$ must be a
critical point of this functional in \emph{all} coordinate charts,
then something geometric, that is, not depending on a particular
choice of coordinate chart, must result. Hence, if we remove the
assumption that the connection is the standard Levi-Civita
connection, then the above set of axioms seems like an interesting place to start.

We note that Einstein and Cartan famously played around with removing the assumption that the connection was
torsion free, while still assuming metric compatibility.  However,
as our beginning point, we make neither assumption. Also, Einstein
and Cartan were not trying to describe dark matter and thus had
different objectives in mind.

\begin{figure}
   \begin{center}
   \includegraphics[height=75mm]{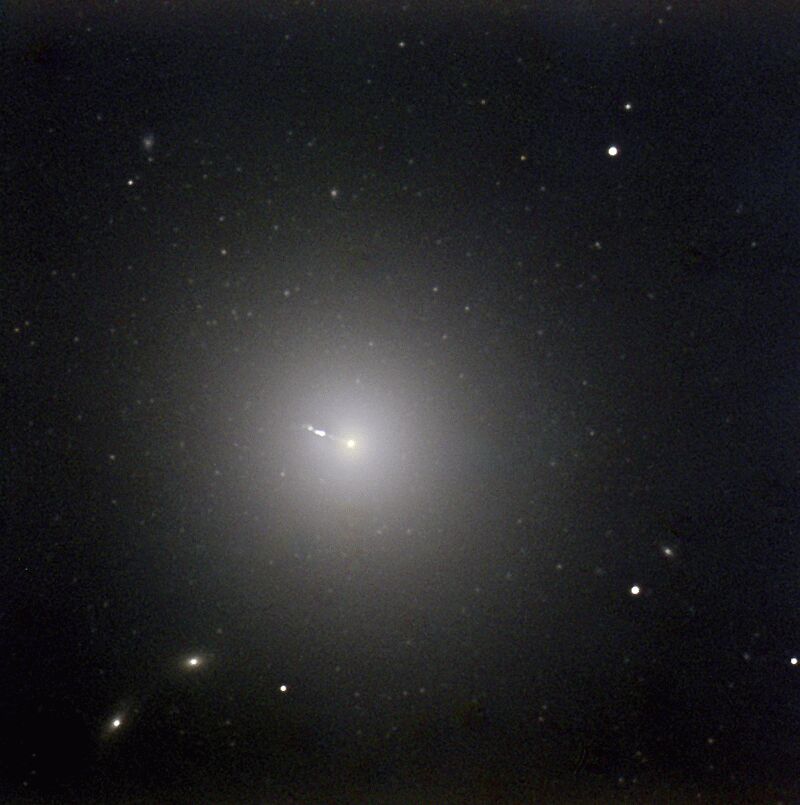}
   \includegraphics[height=75mm]{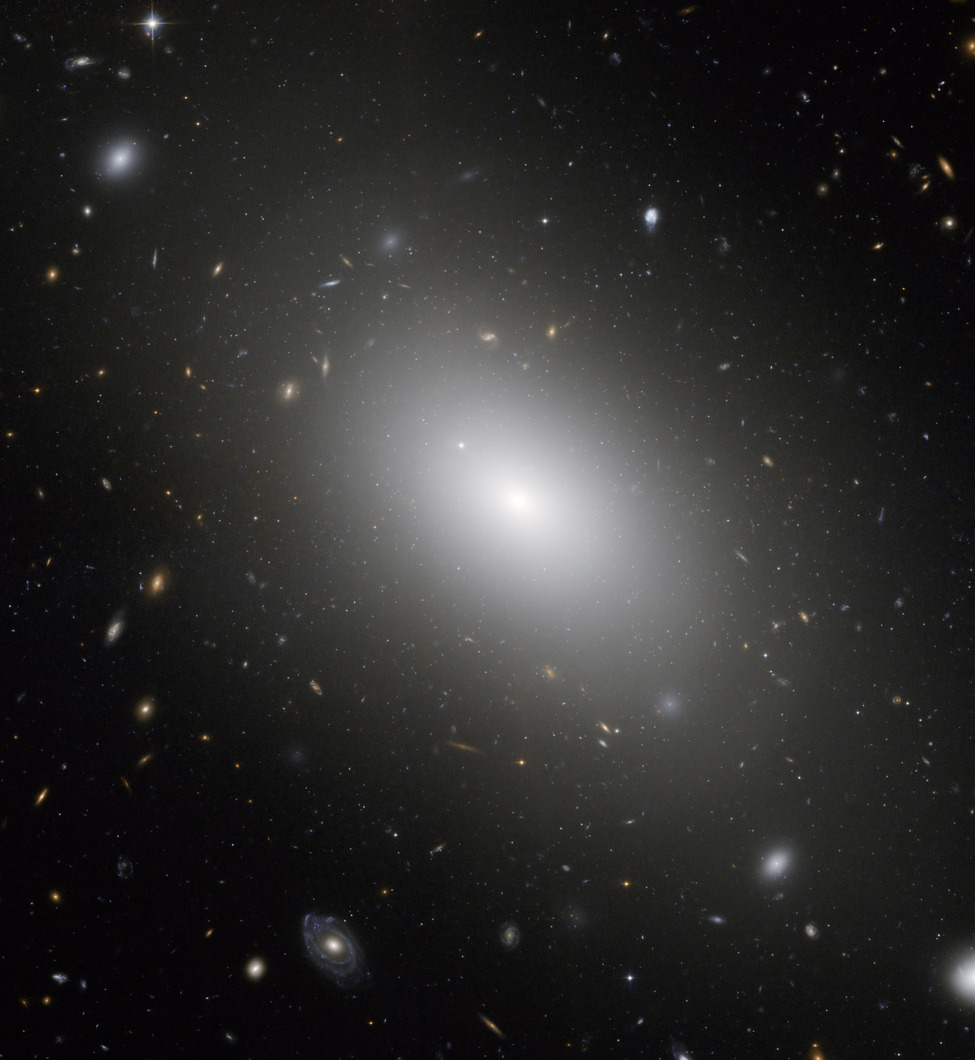}
   \end{center}
   \caption{M 87 (left) and NGC 1132 (right) are typical examples of
   the majority of elliptical galaxies without visible shells.
   Left Image Credit and Copyright:  ING Archive and Nik Szymanek.  Date: 1995.  Telescope:  Jacobus Kapteyn Telescope.  Instrument:  JAG CCD Camera.  Detector:  Tek.  Filters B, V, and R.
   Right Image Credit and Copyright:  NASA, ESA, and the Hubble Heritage (STScI/AURA)-ESA/Hubble
   Collaboration.  Acknowledgment: M. West (ESO, Chile).} \label{F8}
\end{figure}

We refer the reader to the appendices of \cite{DMSG} for a detailed discussion of the implications of Axioms 0 and 1.  The resulting Euler-Lagrange equations for certain actions satisfying these axioms are, quite remarkably, equivalent to the Einstein-Klein-Gordon equations with a cosmological constant $\Lambda$:
\begin{eqnarray}\label{eqn:EE}
   G + \Lambda g &=& 8 \pi \; \left\{\frac{ df \otimes d\bar{f}
                              + d\bar{f} \otimes df}{\Upsilon^2}
   - \left(\frac{|df|^2}{\Upsilon^2} + |f|^2 \right)g \right\} \\
   \Box f &=& \Upsilon^2 f    \label{eqn:KG}
\end{eqnarray}
for a real scalar field $f$, where $f = 0$ everywhere corresponds to the Levi-Civita connection \cite{DMSG} and there is a formula for the connection of the manifold in terms of f and the Levi-Civita connection.  Note that $f = 0$ recovers vacuum general relativity with a cosmological constant.  In the above equations, the speed of light and the universal gravitational constant have been set to one,
$G$ is the Einstein curvature tensor, $\Box$ is the wave operator (the divergence of the gradient) of the Lorentzian metric $g$, $f$ is the scalar field which we suggest causes the curvature of spacetime attributed to dark matter via the right hand side of equation \ref{eqn:EE}, $\Lambda$ is the cosmological constant,
and $\Upsilon$ is another (new) fundamental constant of nature.  Estimates on the value of $\Upsilon$ are computed in \cite{BrayParry} by comparing spherically symmetric solutions of the Einstein-Klein-Gordon equations to data collected about dwarf spheroidal galaxies.

\begin{figure}
   \begin{center}
   \includegraphics[height=59mm]{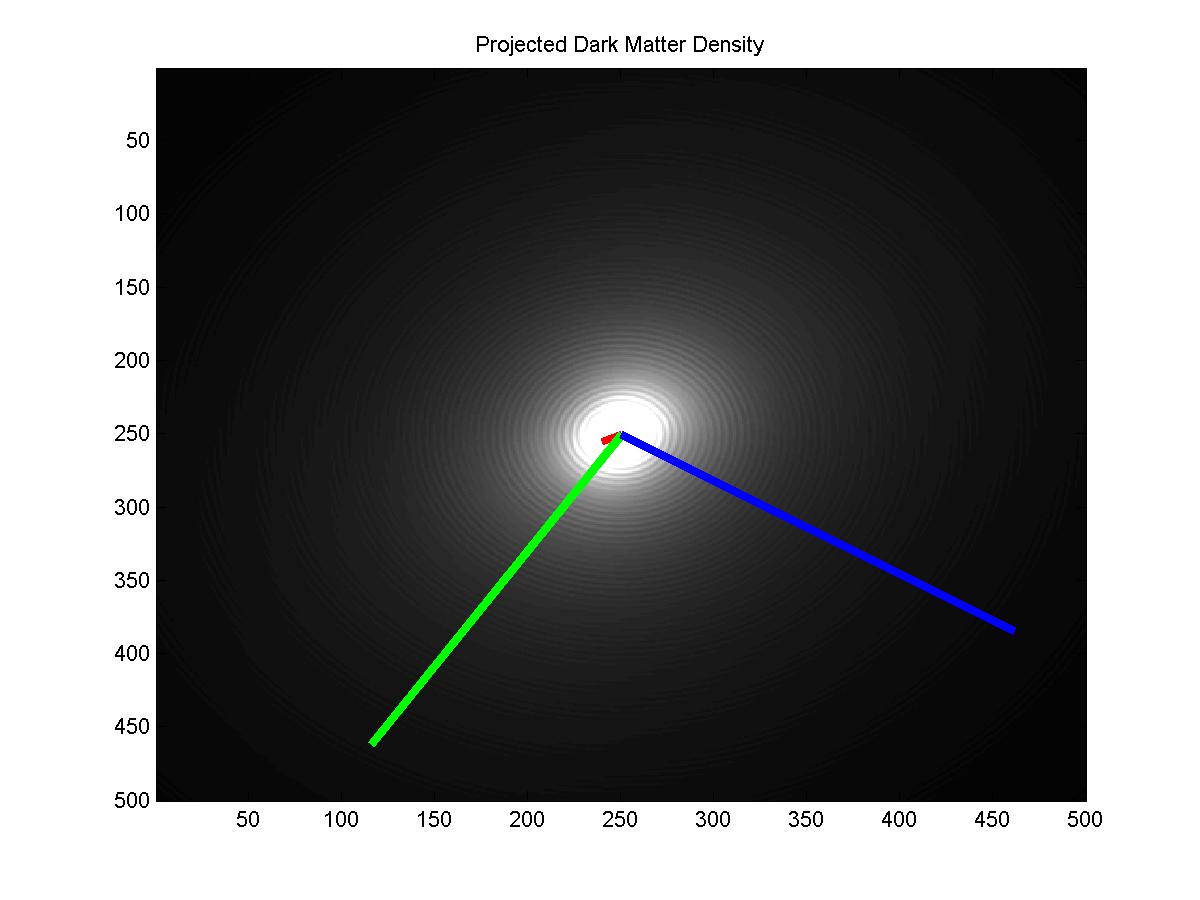}
   \includegraphics[height=59mm]{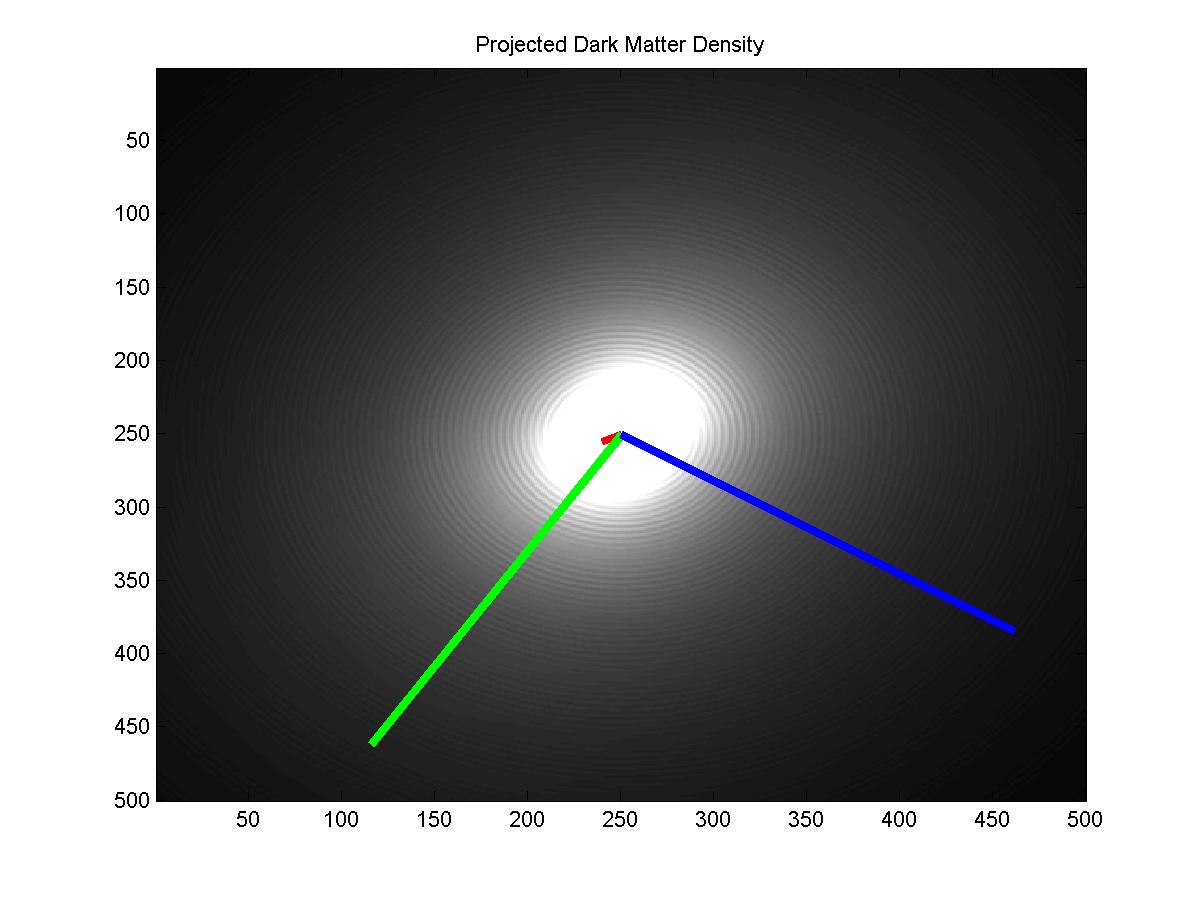}
   \end{center}
   \caption{Simulated images of the same wave dark matter density as in Figure \ref{F2} projected
    into the plane but nearly along the axis of the shells hides the interleaved shells which result from first degree spherical harmonic perturbations from spherically symmetric scalar field solutions to the Klein-Gordon equation.} \label{F9}
\end{figure}

It is more typical for the Klein-Gordon equation to describe a complex scalar field $f$.  If one is willing to allow the connection coefficients $\Gamma_{ijk}$ to be complex valued, then complex valued $f$ are natural as well.  For the Einstein-Klein-Gordon system, a complex scalar field is equivalent to two real scalar fields.
Complex scalar fields are more convenient because they have static spacetime solutions \cite{Parry2,Parry1}.
Analogous ``nearly static'' real solutions can be achieved as $\sqrt2 \, \mbox{Re}(f)$ whose spacetime metrics are very similar except for a high frequency oscillating pressure term which averages out to zero.  Preliminary estimates suggest that this oscillating pressure effect might not be significant enough to make it measurable on physically relevant time scales.  Hence, there may not be a testable difference between the predictions of real and complex scalar field dark matter at this time, though this deserves further thought.  In either case we will call these scalar field models ``wave dark matter.''

Much more is known about the predictions of WIMP dark matter than wave dark matter.  WIMP dark matter has been extremely successful on cosmological scales \cite{WIMPSuccess}.  On the other hand, \cite{DMSG} proves that wave dark matter, unlike WIMP dark matter, is automatically cold, as observed, if the universe is assumed to be homogeneous and isotropic.

There are many basic questions still open about wave dark matter.  Some of the most important questions center around understanding which wave dark matter solutions are physically relevant.  Spherically symmetric static solutions are easy to compute and so are a natural place to start, which raises questions about which of these solutions are stable.
Results in \cite{MSBS} indirectly suggest that regular matter could help stabilize wave dark matter solutions.  Other related results are found in \cite{LaiChoptuik}.  Of course, the most physically relevant wave dark matter solutions could be much more complicated.  Computer simulations of solutions to the Einstein-Klein-Gordon equations with regular matter included as well are needed to gain more insight on these questions of central importance.
 
If wave dark matter is assumed to clump on the scale of galaxies for example, then it would be reasonable to guess (and important to verify) that it has very similar predictions as WIMP dark matter on the cosmological scale.  In \cite{Matos1,Matos2} it is argued that the scalar field dark matter model, which is effectively the same as the wave dark matter model discussed here, should 
reproduce some of the successes of the standard $\Lambda$CDM model
above galactic scales.
Hence, comparing and contrasting the predictions of WIMP dark matter and wave dark matter on the scale of galaxies might be required to find the most obvious differences in their predictions.

\subsection{Dark Matter on the Galactic Scale}

In this subsection we list six conclusions about dark matter based on astronomical observations.  We have roughly ordered the list by confidence in the statement and have placed parentheses around the statements which are less certain.

%\newpage
\begin{center}
{\bf Conclusions from Astronomical Observations}
\end{center}

\begin{figure}
   \begin{center}
   \includegraphics[height=75mm]{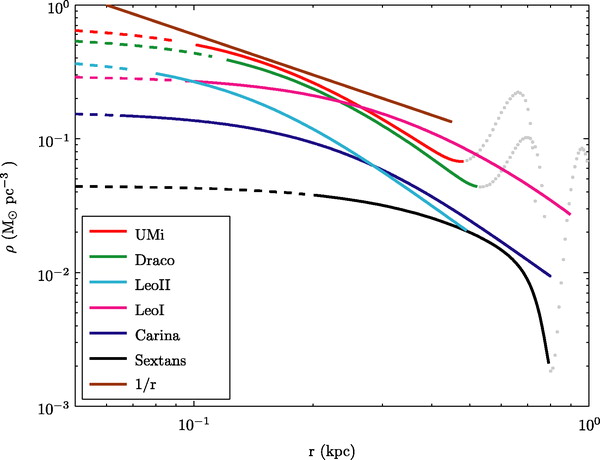}
   \end{center}
   \caption{Density as a function of radius for six dwarf spheroidal galaxies based on the distribution of the velocities of the stars in each galaxy \cite{Gilmore}.  Note that three densities have oscillations in their density profiles (arbitrarily dashed in gray starting at the first local minimum).  Given the uncertainties in the data and the assumptions of the model, it is unclear whether or not these oscillations in the dark matter density actually exist.}\label{GilmorePlot}
\end{figure}

\begin{enumerate}
  \item Disk galaxies commonly have spiral and barred spiral density wave patterns in their images \cite{BM, BT}.\label{I1}
  \item Elliptical galaxies commonly have interleaved shells in their images \cite{BM, BT}.\label{I2}
  \item Most of the dark matter in the universe is cold (traveling at non-relativistic speeds) \cite{BM,BT}.\label{I3}
  \item (There is a rough lower bound on the mass of isolated ``blobs'' of dark matter \cite{MissingSatellitesProblem,DMAW}.)  \label{I4}
  \item ((The dark matter density in the cores of galaxies appears to be bounded \cite{DMAW, LiChen, Flat, CoreCusp}.))\label{I5}
  \item (((It is possible that there are oscillations in the dark matter density as a function of radius, as in Figure \ref{GilmorePlot}, for some dwarf spheroidal galaxies.)))\label{I6}
\end{enumerate}

There are many other known facts about dark matter in galaxies that we have not listed here \cite{DMAW}.  For example, the rotation curves of spiral galaxies provide a tremendous amount of information about the distribution of dark matter in spiral galaxies and suggest the
notion of a universal rotation curve \cite{DMAW}.  Also, the Tully-Fisher relation \cite{McGaugh2011,McGaugh2009}, that the luminous mass of a spiral galaxy divided by the typical velocity of stars in the galaxy to the 4th power is roughly a constant for every spiral galaxy in the known universe, is a very remarkable fact deserving of a great deal of further study and consideration.  Either of these last two observations could be extremely important for revealing the true nature of dark matter.

For now, though, we focus on the six conclusions listed above because plausible statements can be made differentiating wave dark matter and WIMP dark matter with regards to these conclusions based on astronomical observations. The first two conclusions are observations and hence are beyond dispute.  The third conclusion is for all intents and purposes beyond dispute given that dark matter appears to be gravitationally trapped in galaxies and clusters of galaxies, which would not be the case if it were moving at speeds close to the speed of light.

The last three conclusions, however, are somewhere between probable and possible.  Naturally it is frustrating not to have complete clarity on important issues, but given that dark matter is invisible and known only by its gravity, we have to take what we can get.

The smallest known ``blobs'' of dark matter are dwarf speroidal galaxies.  In addition, dwarf spheroidal galaxies are almost entirely dark matter \cite{DMAW}, which is one reason they may be one of the best sources of information about dark matter.  Another curious fact is that there appears to be a rough lower bound on the masses of dwarf spheroidal galaxies, below which ``blobs'' of dark matter have not been observed to exist  \cite{MissingSatellitesProblem,DMAW}.  By rough, we include the possibility that the frequency of such blobs of dark matter drops off quickly as a function of total mass at some point.  On the other hand, we have put conclusion \ref{I4} in parentheses since dark matter is, after all, invisible.  Hence, maybe smaller blobs of dark matter exist, but we just can not see them.  This issue is commonly referred to as ``The Missing Satellites Problem'' \cite{MissingSatellitesProblem}.

When WIMP dark matter is modeled on computers under a certain range of initial conditions, it is found to settle down to a density distribution which is well approximated by the spherically symmetric Navarro-Frenk-White profile given by
\begin{equation}\label{eqn:NFW}
   \rho(r) = \frac{\rho_0}{\frac{r}{R_s}\left(1 + \frac{r}{R_s}\right)^2}
\end{equation}
for some given values of $\rho_0$ and the scale radius $R_s$ \cite{NFW1, NFW2, CoreCusp}.  In particular, the density at the origin (assumed to be the center of mass of the system) diverges to infinity in this density profile like $\rho_0 R_s/r$.  Given this, astronomers have looked for evidence for the dark matter density in galaxies to be unbounded at their centers. However, so far no such density spikes in dark matter have been found, which is the content of conclusion \ref{I5}, sometimes referred to as ``The Core-Cusp Problem'' \cite{CoreCusp}.

The weaknesses of conclusions \ref{I4} and \ref{I5} is that they concern ``not observing'' something as opposed to directly observing something.  We have ranked conclusion \ref{I4} higher than conclusion \ref{I5} because masses of galaxies are found in a wide continuum right down to the lower limit at which point there is close to nothing.  Hence, the actual observation of galaxies with masses in the entire range above the cutoff is very relevant.  Nevertheless, concerning Conclusion \ref{I5}, astronomers have had better success fitting what they call ``cored'' density profiles (Navarro-Frenk-White with the spike removed, like coring an apple) to dark matter densities deduced in galaxies from observations.

Finally, conclusion \ref{I6} is the least certain conclusion of the bunch and falls into the ``possible'' category.  More observations and analyses are required to determine whether observation \ref{I6} is really true, so the reader is duly cautioned.  Nevertheless, conclusion \ref{I6}, if true, could have far reaching implications about the nature of dark matter, so this author decided it deserved mentioning.

Conclusion \ref{I6} is based entirely on the work in \cite{Gilmore}, though those authors do not claim this conclusion in their paper.  In \cite{Gilmore}, the authors attempt to compute the spherically symmetric density as a function of radius of six dwarf spheroidal galaxies based on the observed velocities of the stars in each galaxy.  This type of inverse calculation has a list of assumptions built into it.  In addition, there are error bars in such calculations coming from uncertainties in the data which were not clear to this author from reading the paper. Nevertheless, the result of the model used in \cite{Gilmore} is that half of the six dwarf spheroidal galaxies had oscillations in their densities as a function of radius in order to best fit the data, as seen in Figure \ref{GilmorePlot}.  Given the uncertainties in the data and the assumptions of the model, it is unclear to this author whether or not these oscillations in the dark matter density in these three dwarf spheroidal galaxies can actually be concluded to exist.  However, even if the error bars are such that these oscillations may not be concluded to exist, one may conclude that oscillations in dark matter density as a function of radius are at least compatible with these particular observations.

In fact, in \cite{Gilmore}, the authors suggest these dark matter density oscillations are not real, though there is very little discussion.  Given the pervasiveness of the WIMP model of dark matter and the fact that these oscillations are highly unexpected from a WIMP dark matter point of view, this author was concerned that the possibility of dark matter density oscillations in dwarf spheroidal galaxies may have been discarded prematurely.  Hence, we have included conclusion \ref{I6}, perhaps better named ``possibility \ref{I6},'' here to stimulate more discussion of this provocative possibility.

\subsection{Wave Dark Matter on the Galactic Scale}

So how does wave dark matter compare to WIMP dark matter in terms of predictions each makes on the galactic scale?  Our discussion here, while relevant, does not come close to settling this complex question.

In this subsection, we discuss the ways in which
the six ``Conclusions from Astronomical Observations'' about dark matter in galaxies from the previous subsection favor the possibility of wave dark matter as a viable dark matter model.
We believe that the discussion presented here justifies wave dark matter as a legitimate dark matter candidate well deserving of further study.

\newcounter{saveenum}
\begin{enumerate}
  {\bf \item Disk galaxies commonly have spiral and barred spiral density wave patterns in their images \cite{BM, BT}.}\label{II1}
  \setcounter{saveenum}{\value{enumi}}
\end{enumerate}

\begin{enumerate}
  \setcounter{enumi}{\value{saveenum}}
{\bf   \item Elliptical galaxies commonly have interleaved shells in their images \cite{BM, BT}.}\label{II2}
  \setcounter{saveenum}{\value{enumi}}
\end{enumerate}

Even though most of the mass of galaxies is dark matter, in the standard explanations of these two phenomena dark matter plays a secondary role.  If this is true, wave dark matter and WIMP dark matter are both able to fit this picture since both have spherically symmetric solutions.  However, looking at first and second degree spherical harmonic perturbations from spherical symmetry for wave dark matter solutions gives an additional possible mechanism by which to explain each of these phenomena.  In these wave dark matter mechanisms, the substructure of the dark matter (the perturbations from spherical symmetry) is what drives the substructure in the regular matter (spiral patterns and shells).

In the standard Lin-Shu density wave theory, spiral galaxies are modeled with spiral density waves resulting from global gravitational instabilities in the visible matter.  The dark matter is modeled as spherically symmetric and any dark matter substructure is assumed to be of secondary importance.  By contract, in \cite{DMSG} we show how second degree spherical harmonic deviations from spherical symmetry in the wave dark matter can lead to rotating ellipsoidal galactic potentials which can produce spiral patterns in the gas, dust, and stars, thereby accounting for both spiral and barred spiral patterns in disk galaxies, as demonstrated in Figures \ref{SG1}, \ref{SG2}, \ref{SG3}, and \ref{SG4}.

Similarly, the standard explanations for the observed interleaved shells in elliptical galaxies as summarized on pages 202-203 of \cite{BM} also model dark matter as being roughly spherically symmetric and make no reference to dark matter substructure.  By contrast, in this paper we show how first degree spherical harmonic deviations from spherical symmetry in the wave dark matter leads to interleaved shells in the dark matter density, as shown in Figure \ref{F2}.  As we suggested in this paper already, this global interleaved shell structure in the dark matter, while invisible itself, could conceivably play an important role in the formation of visible interleaved shells of regular matter through gravity, friction, dynamical friction, and other processes, though this is a complex open question.

Note that the perturbations in the wave dark matter solutions that we are suggesting are first degree spherical harmonic perturbations to explain shells in elliptical galaxies and second degree spherical harmonic perturbations to explain spiral and barred spiral patterns in disk galaxies.  Our reasoning for this difference goes as follows:  Typically, we would expect first degree spherical harmonic perturbations to be larger than second degree spherical harmonic perturbations which, in turn, we would expect to be larger than third degree spherical harmonic perturbations, etc.  This picture is compatible with our modeling of shells in elliptical galaxies, but raises the question of why we assume second degree spherical harmonic perturbations (instead of first degree) typically dominate is spiral galaxies.

Our reasoning is that first degree spherical harmonic perturbations of wave dark matter are the only ones which do not fix the center of mass of the wave dark matter at the origin.  Since the overall center of mass is fixed at the origin, this means that the regular matter center of mass would have to oscillate as well.  We speculate that spiral galaxies have a roughly fixed center of mass for the regular matter, perhaps due to the important role played by friction of the gas and the dust in the galaxy.  Hence, a roughly fixed center of mass for the regular matter in a spiral galaxy would not allow for a first degree spherical harmonic perturbation to dominate, in which case we would expect a second degree spherical harmonic perturbation of the wave dark matter to dominate.

While wave dark matter substructure offers tantalizing possibilities to explain visibile matter substructure in the images of both spiral and elliptical galaxies, much more study is warranted.  Detailed computer simulations which model the relevant complicated astrophysical processes correctly are mostly likely required to make substantial further progress.

\begin{enumerate}
  \setcounter{enumi}{\value{saveenum}}
  {\bf \item Most of the dark matter in the universe is cold (traveling at non-relativistic speeds) \cite{BM,BT}.}\label{II3}
  \setcounter{saveenum}{\value{enumi}}
\end{enumerate}

Unlike WIMP dark matter, wave dark matter must be cold in a homogeneous isotropic universe \cite{DMSG}.  Since dark matter is observed to be cold, it is reasonable to call this a successful prediction of wave dark matter. Had the dark matter in the universe been observed to be hot or warm and homogeneous and isotropic, this would have ruled out wave dark matter as a reasonable dark matter model.

The difference in this regard between wave dark matter and WIMP dark matter has to do with the wave nature of wave dark matter and the particle nature of WIMP dark matter.  In WIMP dark matter, the WIMPs can be taken to be uniformly spread over the universe at every point and going in every direction at {\it any} fixed velocity one likes.  Hence, homogeneous isotropic universes can have WIMP dark matter at any temperature, from zero velocity to close to the speed of light.  With wave dark matter, however, the stress-energy tensor is not additive in this same way.  The wave function describing the wave dark matter is additive, but the stress-energy tensor is quadratic in the wave function.  As shown in \cite{DMSG}, homogeneity and isotropy of the universe imply that the wave dark matter is very cold at every point as long as the rate of change of the Hubble constant is much smaller in norm than $\Upsilon^2$, the constant from the Klein-Gordon equation.  More precisely, in Appendix D of \cite{DMSG}, we prove the following theorem.

\begin{theorem}
Suppose that the spacetime metric is both homogeneous and isotropic,
and hence is the Friedmann-Lema\^{i}tre-Robertson-Walker metric
$-dt^2 + a(t)^2 ds^2$, where $ds^2$ is any constant curvature metric.
If $f(t,\vec{x})$ is a
real-valued solution to the Klein-Gordon equation
$\Box_g f = \Upsilon^2 f$
with a stress-energy tensor
which is isotropic, then $f$ is solely a function of $t$.
Furthermore, if we let $H(t) = a'(t)/a(t)$ be the Hubble constant
(which of course is actually a function of $t$), and $\rho(t)$ and
$P(t)$ be the energy density and pressure of the scalar field at
each point, then
\begin{equation}
   \frac{\bar{P}}{\bar{\rho}} = \frac{\epsilon}{1 + \epsilon}
\end{equation}
where
\begin{equation}
   \epsilon = - \frac{3\overline{H'}}{4\Upsilon^2}
\end{equation}
and
\begin{equation}
   \bar{\rho} = \frac{1}{b-a} \int_a^b \rho(t) \;dt , \hspace{.2in}
   \bar{P} = \frac{1}{b-a} \int_a^b P(t) \;dt, \hspace{.2in}
   \overline{H'} = \frac{\int_a^b H'(t) f(t)^2 \;dt}
   {\int_a^b f(t)^2 \;dt} ,
\end{equation}
where $a,b$ are two zeros of $f$ (for example, two consecutive
zeros).
\end{theorem}

In this theorem, the temperature of the universe is approximated by the quantity $\frac{\bar{P}}{\bar{\rho}}$. Hence, if the rate of change of the Hubble constant $H'(t)$ is much smaller in norm than $\Upsilon^2$, then $\overline{H'}$ must be as well.  Hence, $\epsilon$ and the temperature expression $\frac{\bar{P}}{\bar{\rho}}$ must be close to zero.  This is the precise sense in which homogeneity and isotropy of the universe imply that wave dark matter must be cold.

Of course the universe is not perfectly homogeneous and isotropic, so it is an interesting direction to take this theorem and try to study what these types of arguments imply when the universe is only assumed to be approximately homogeneous and isotropic.  Also, a similar result is true when the scalar field is allowed to be complex since one complex scalar field is equivalent to two real scalar fields.

\begin{enumerate}
  \setcounter{enumi}{\value{saveenum}}
{\bf  \item (There is a rough lower bound on the mass of isolated ``blobs'' of dark matter \cite{DMAW}.)  } \label{II4}
\setcounter{saveenum}{\value{enumi}}
\end{enumerate}

WIMP dark matter simulations predict clumps of dark matter on all scales, but this is not what is observed \cite{MissingSatellitesProblem}.  Instead, clumps of dark matter have been observed on all scales down to a smallest one, the mass of dwarf spheroidal galaxies, and then nothing more \cite{DMAW}.

It is not definitively clear what wave dark matter predicts in this regard.  However, static spherically symmetric solutions to the Einstein-Klein-Gordon equations which describe wave dark matter come in discrete steps \cite{Parry2} as a ground state, first excited state, second excited state, etc.  Each of these ``solutions'' is actually a one parameter family of solutions which allows the total mass of the clump of wave dark matter to vary.  However, if for some reason there is a natural choice of scale for each of these excited states, perhaps given by a preferred time frequency for the wave dark matter, then this would predict a lowest mass size for a blob of wave dark matter given by the ground state solution.  Understanding wave dark matter in this regard is a very good direction to pursue.

\begin{enumerate}
  \setcounter{enumi}{\value{saveenum}}
{\bf  \item ((The dark matter density in the cores of galaxies appears to be bounded \cite{DMAW, LiChen, Flat, CoreCusp}.))}\label{II5}
  \setcounter{saveenum}{\value{enumi}}
\end{enumerate}

WIMP dark matter, as discussed in the previous subsection, predicts a spike in the dark matter density going to infinity at the centers of galaxies \cite{CoreCusp, NFW1, NFW2}.  Wave dark matter, on the other hand, does not.  Detecting such a dark matter spike would most likely rule out wave dark matter as a reasonable dark matter candidate.  However, as such spikes have not been observed yet, and as astronomers are leaning towards the conclusion that they are not there \cite{DMAW, LiChen, Flat}, this observation, or lack of one, perhaps favors wave dark matter at this time.  This issue is important for astronomers to continue to study as detection of a spike in the dark matter density or a definitive demonstration that such a spike does not exist has important implications for the nature of dark matter.

\begin{enumerate}
   \setcounter{enumi}{\value{saveenum}}
   {\bf \item (((It is possible that there are oscillations in the dark matter density as a function of radius for some dwarf spheroidal galaxies.)))}\label{II6}
\end{enumerate}

As already discussed in the previous subsection, it is not clear whether this possibility is actually true, so gathering more data and studying the existing data for dwarf spheroidal galaxies is warranted.  However, qualitatively speaking, oscillations in the dark matter density as a function of radius for some dwarf spheroidal galaxies is a very reasonable prediction from the point of view of wave dark matter since the static spherically symmetric solutions for wave dark matter all have this feature, except for the ground state which has monotonically decreasing density as a function of radius \cite{Parry2}.  The radial density function for WIMP dark matter, on the other hand, would be expected to be closer to the monotonically decreasing Navarro-Frenk-White profile in equation \ref{eqn:NFW}.  Perhaps our main contribution here is to ask astronomers and astrophysicists to continue to try to estimate dark matter density in the smallest galaxies, or perhaps those a little bit bigger than the smallest galaxies, to see if oscillations in the dark matter density as a function of radius are there or not.  It is also a good problem to try to figure out what the precise predictions of the two dark matter theories are as well.

In summary, these six conclusions about dark matter based on observations show that there are many questions that remain about the true nature of dark matter.  As such, both WIMP dark matter and wave dark matter deserve much further study to see how their predictions compare to observations, especially on the galactic scale.

\begin{figure}
   \begin{center}
   \includegraphics[width=160mm]{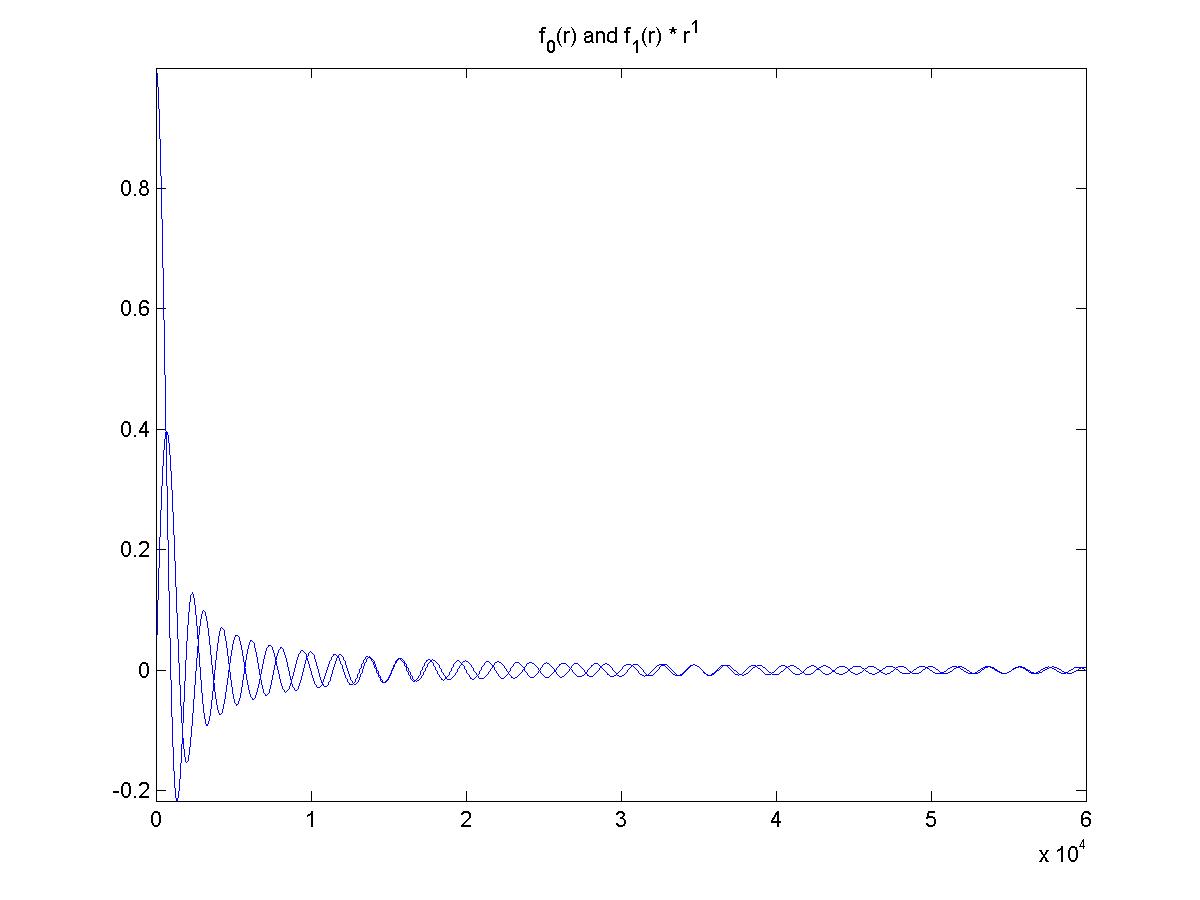}
   \end{center}
   \caption{Plots of $f_{\omega_0,0}(r)$ and $r \cdot f_{\omega_1,1}(r)$ defined by equation \ref{sphericalode} and essential for understanding the wave dark matter density in equation \ref{dmd}.  The fact that these two oscillating functions go in and out of phase is what causes the interleaved shell patterns in the wave dark matter density.  Image created by running the matlab command shells(500, .1, 100000, 60000, 1, -1, 5, 10), available at the author's ``Wave Dark Matter Web Page'' at http://www.math.duke.edu/\texttildelow bray/darkmatter/darkmatter.html. } \label{inandoutofphase}
\end{figure}

\begin{figure}
   \begin{center}
   \includegraphics[width=160mm]{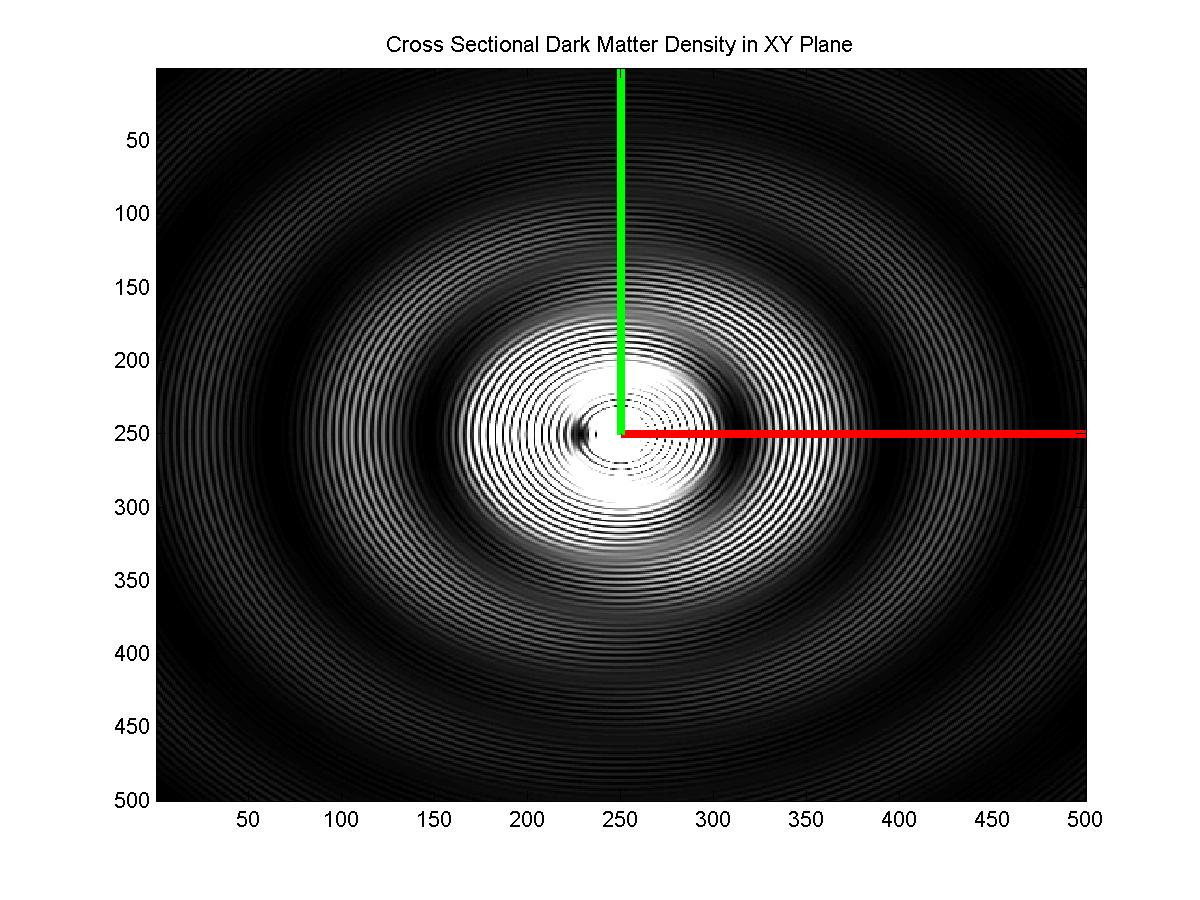}
   \end{center}
   \caption{Cross sectional wave dark matter density in the $xy$ plane.  The red axis is the $x$-axis and the green axis is the $y$-axis.  Image created by running the matlab command shells(500, .1, 100000, 60000, 1, -1, 5, 10), available at the author's ``Wave Dark Matter Web Page'' at http://www.math.duke.edu/\texttildelow bray/darkmatter/darkmatter.html.} \label{CS1}
\end{figure}

\begin{figure}
   \begin{center}
   \includegraphics[width=160mm]{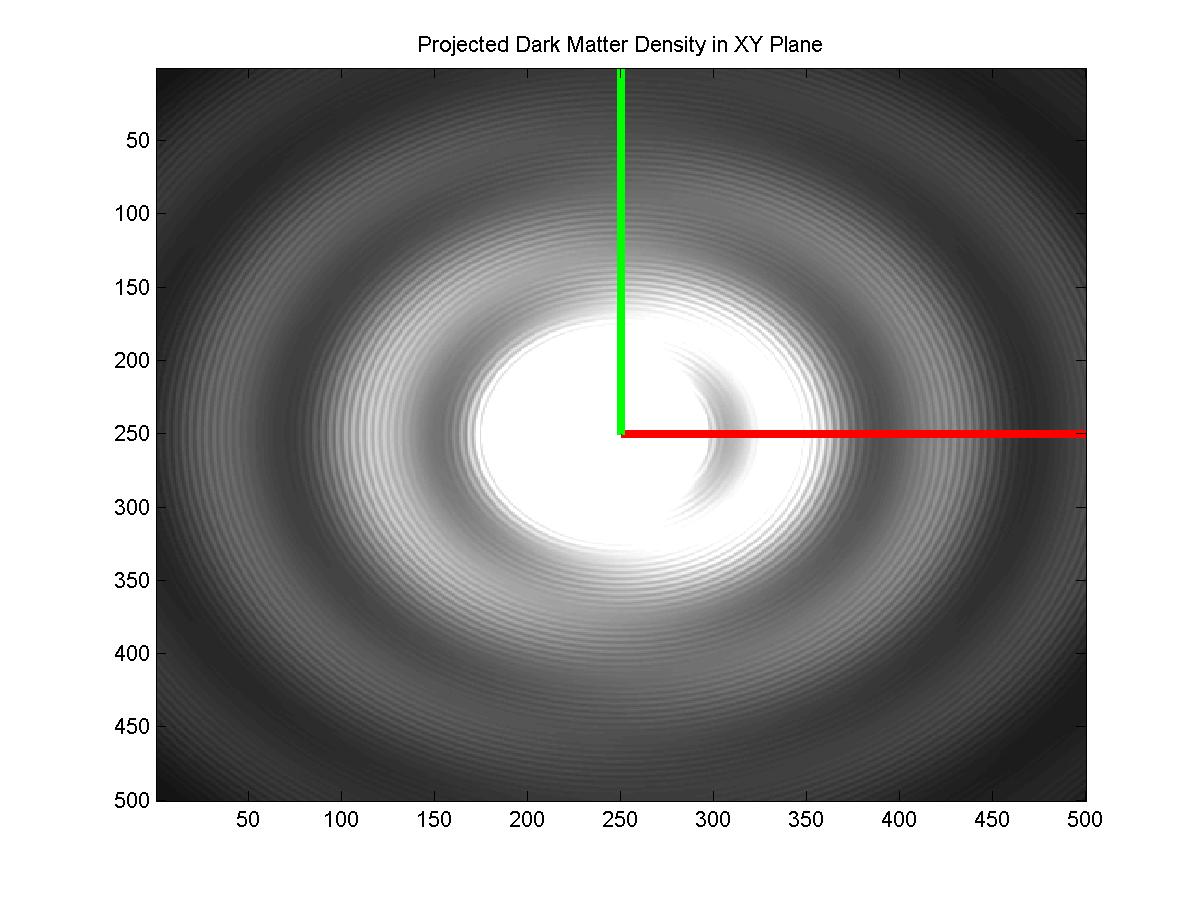}
   \end{center}
   \caption{Projected wave dark matter density in the $xy$ plane.  The red axis is the $x$-axis and the green axis is the $y$-axis.  This is what the wave dark matter would look like if it glowed white, as observed from a distant observer looking down from the point of view of the $z$-axis.  Image created by running the matlab command shells(500, .1, 100000, 60000, 1, -1, 5, 10), available at the author's ``Wave Dark Matter Web Page'' at http://www.math.duke.edu/\texttildelow bray/darkmatter/darkmatter.html.} \label{P1}
\end{figure}

\begin{figure}
   \begin{center}
   \includegraphics[width=160mm]{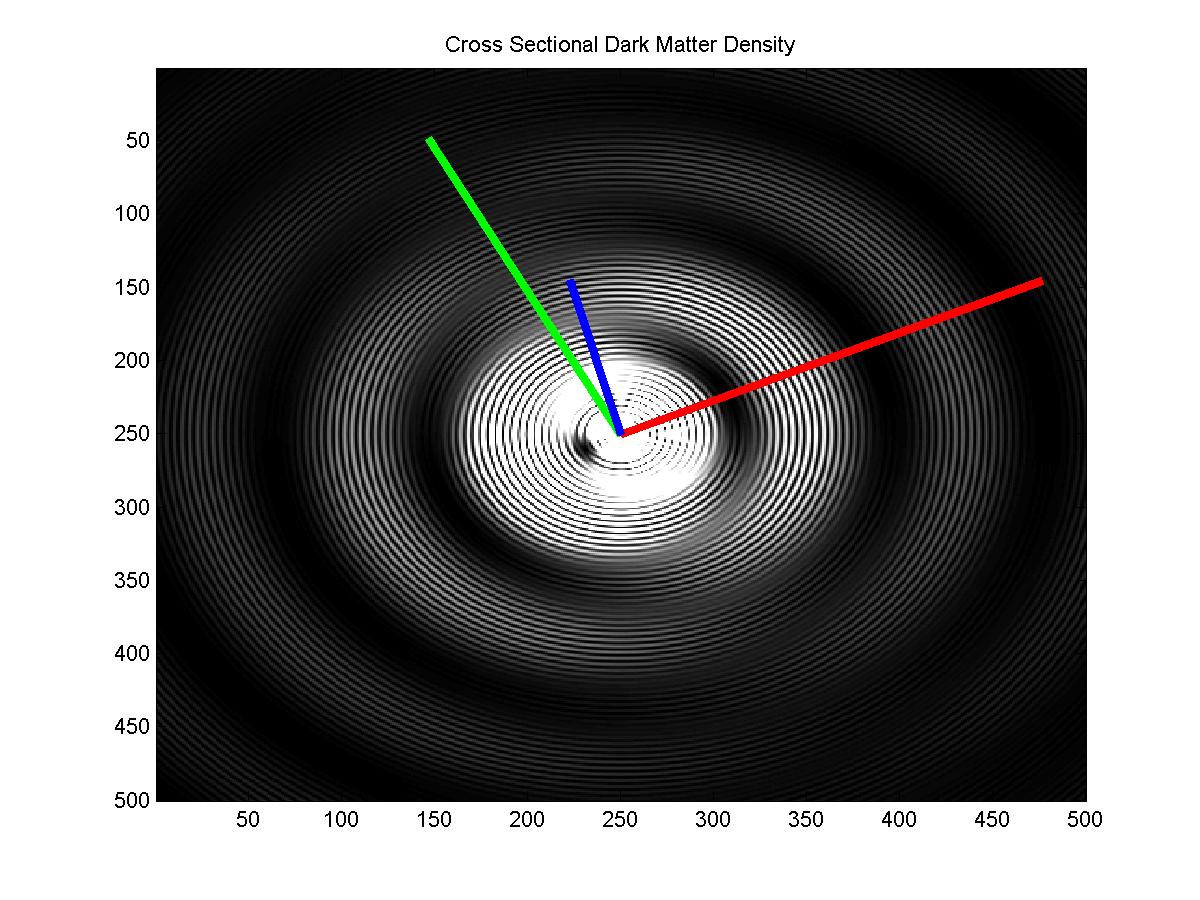}
   \end{center}
   \caption{Cross sectional wave dark matter density in a random plane, where the $x$-axis is red, the $y$-axis is green, and the $z$-axis is blue.  Image created by running the matlab command shells(500, .1, 100000, 60000, 1, -1, 5, 10), available at the author's ``Wave Dark Matter Web Page'' at http://www.math.duke.edu/\texttildelow bray/darkmatter/darkmatter.html.} \label{CS2}
\end{figure}

\begin{figure}
   \begin{center}
   \includegraphics[width=160mm]{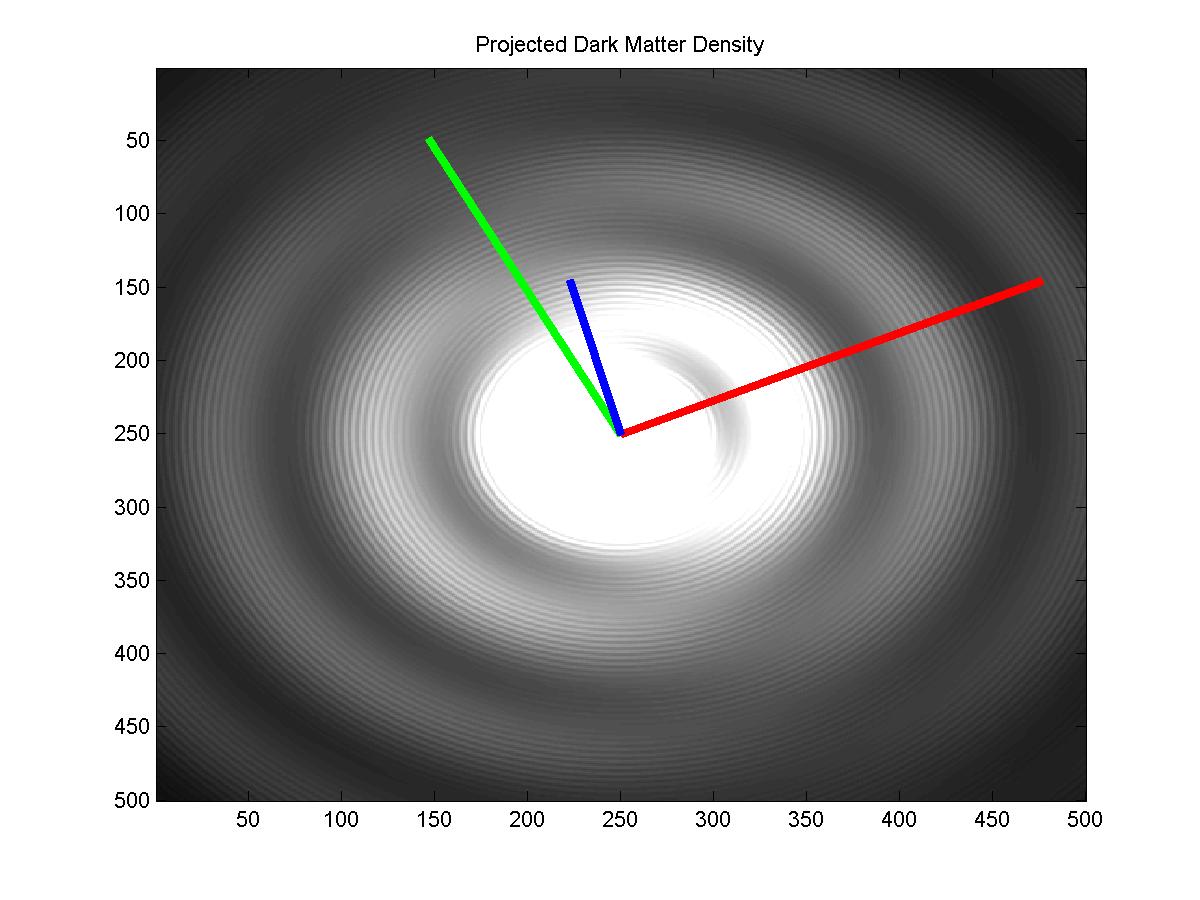}
   \end{center}
   \caption{Projected wave dark matter density in the same plane as the previous figure, where the $x$-axis is red, the $y$-axis is green, and the $z$-axis is blue.  This is what the wave dark matter would look like if it glowed white.  Image created by running the matlab command shells(500, .1, 100000, 60000, 1, -1, 5, 10), available at the author's ``Wave Dark Matter Web Page'' at http://www.math.duke.edu/\texttildelow bray/darkmatter/darkmatter.html.} \label{P2}
\end{figure}

\begin{figure}
   \begin{center}
   \includegraphics[width=160mm]{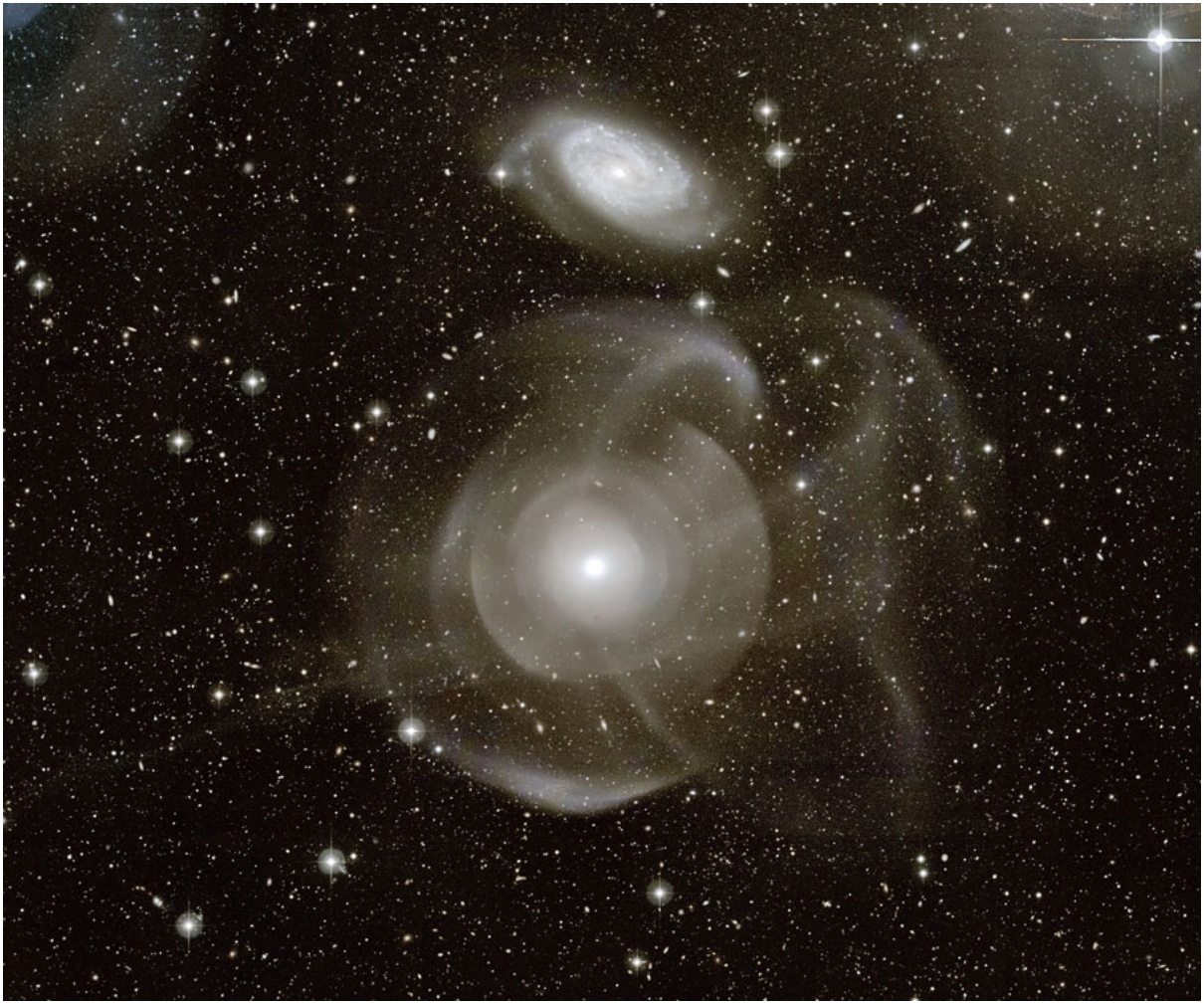}
   \end{center}
   \caption{Shells in Elliptical Galaxies:   NGC 474 is an example of an
   elliptical galaxy with shells.  Image Credit and Copyright: P.-A. Duc (CEA, CFHT), Atlas 3D Collaboration.  From http://apod.nasa.gov/apod/ap110726.html, the NASA Astronomy Picture of the Day: ``The multiple layers of emission appear strangely complex and unexpected given the relatively featureless appearance of the elliptical galaxy in less deep images. The cause of the shells is currently unknown, but possibly tidal tails related to debris left over from absorbing numerous small galaxies in the past billion years. Alternatively the shells may be like ripples in a pond, where the ongoing collision with the spiral galaxy just above NGC 474 is causing density waves to ripple though the galactic giant. Regardless of the actual cause, the above image dramatically highlights the increasing consensus that at least some elliptical galaxies have formed in the recent past, and that the outer halos of most large galaxies are not really smooth but have complexities induced by frequent interactions with -- and accretions of -- smaller nearby galaxies. The halo of our own Milky Way Galaxy is one example of such unexpected complexity. NGC 474 spans about 250,000 light years and lies about 100 million light years distant toward the constellation of the Fish (Pisces).''} \label{Figure:NGC474}
\end{figure}

\begin{figure}
   \begin{center}
   \includegraphics[width=50mm]{f132_500_0p1_100000_60000_1_-1_5_10_1904p7619_2105p2632.jpg}
   \includegraphics[width=50mm]{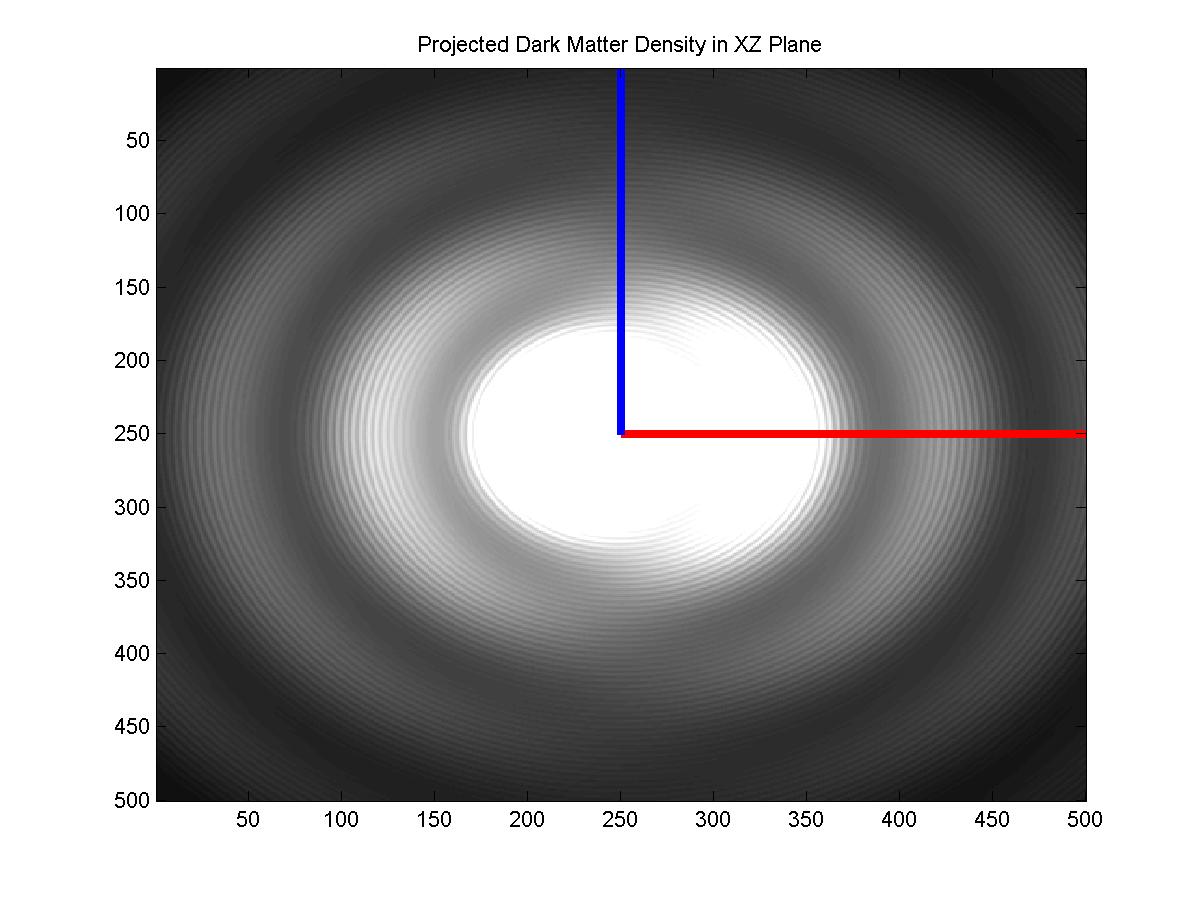}
   \includegraphics[width=50mm]{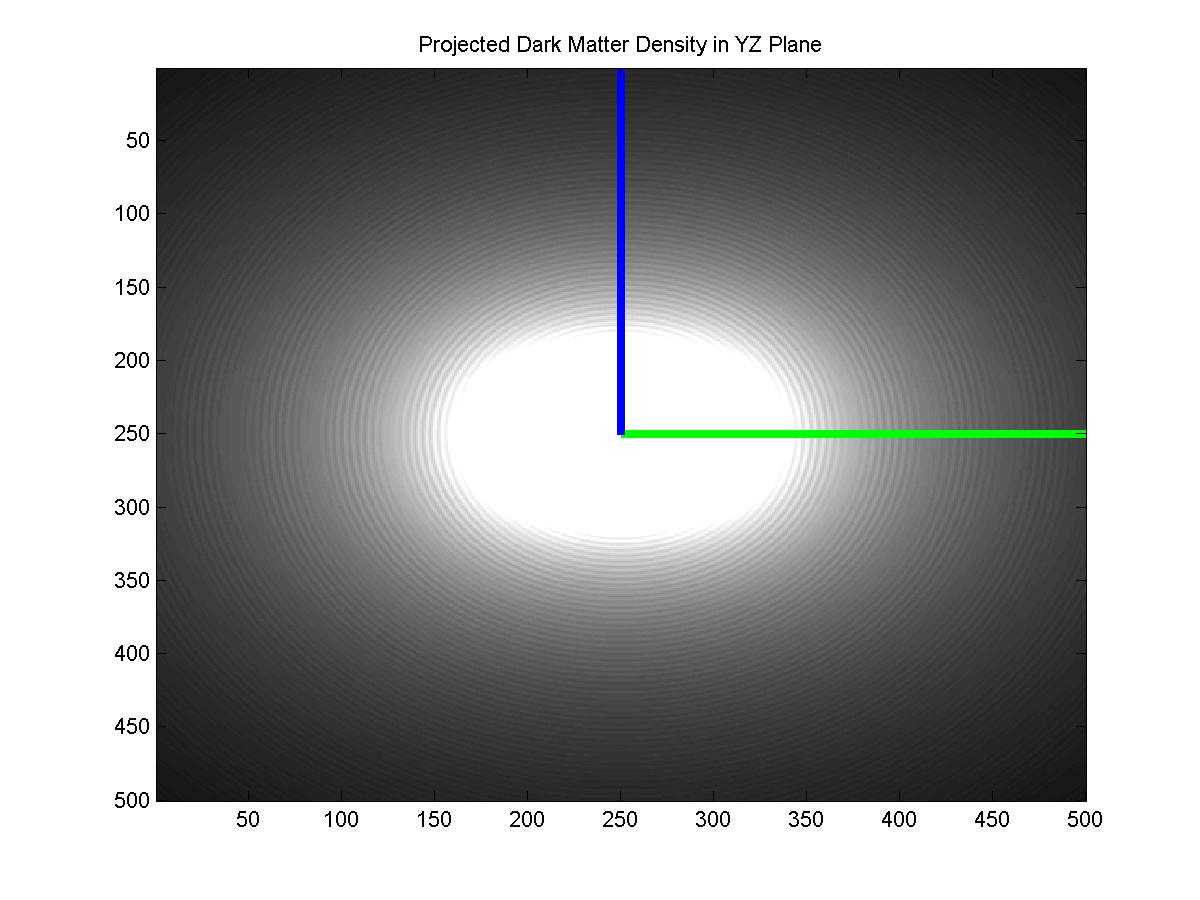}
   \includegraphics[width=50mm]{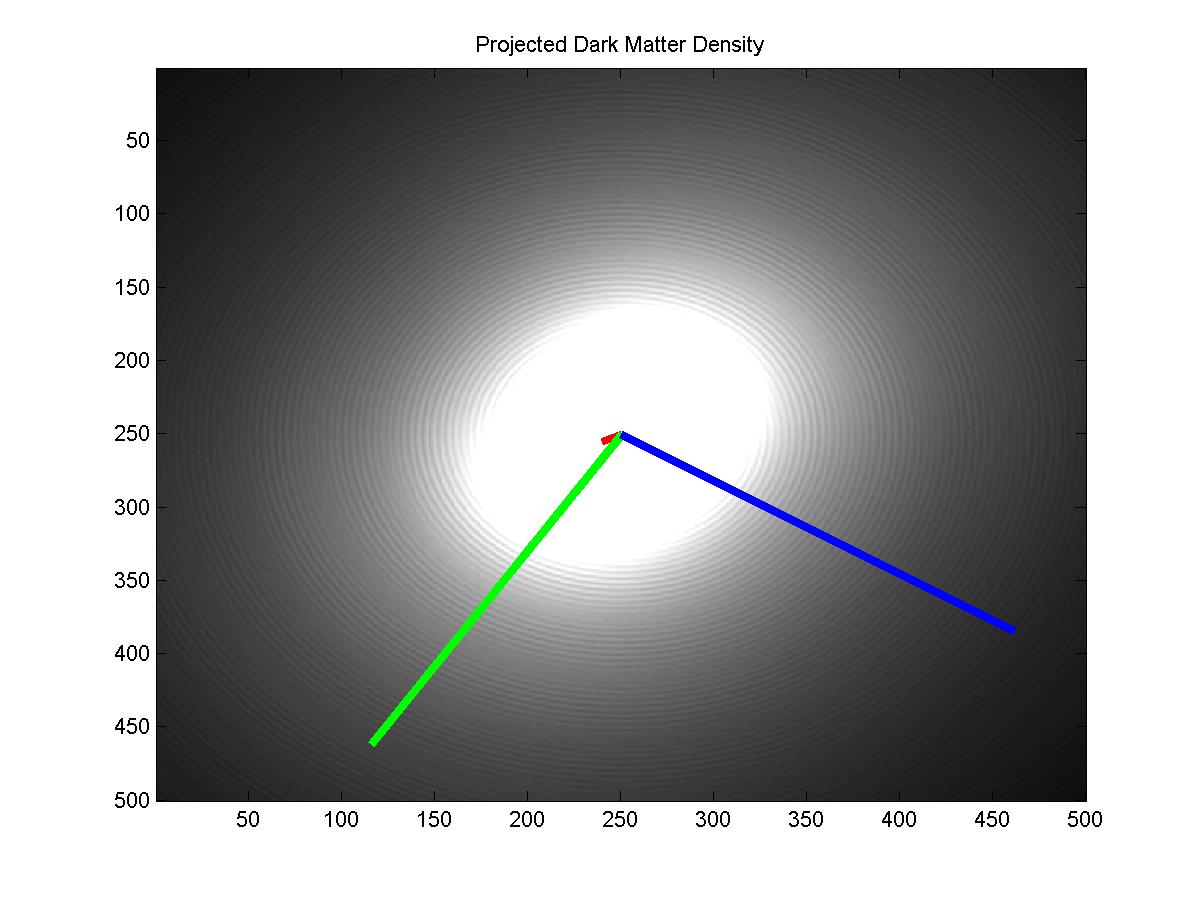}
   \includegraphics[width=50mm]{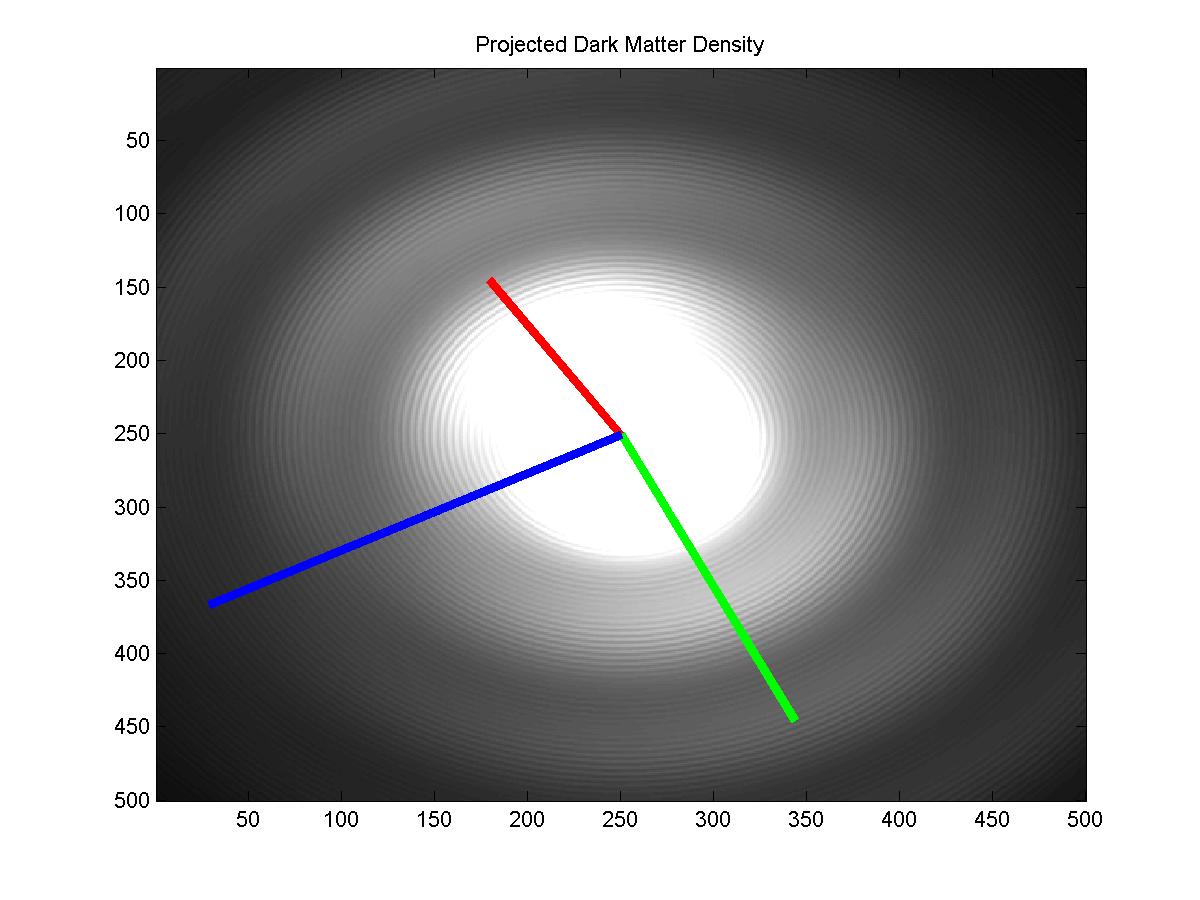}
   \includegraphics[width=50mm]{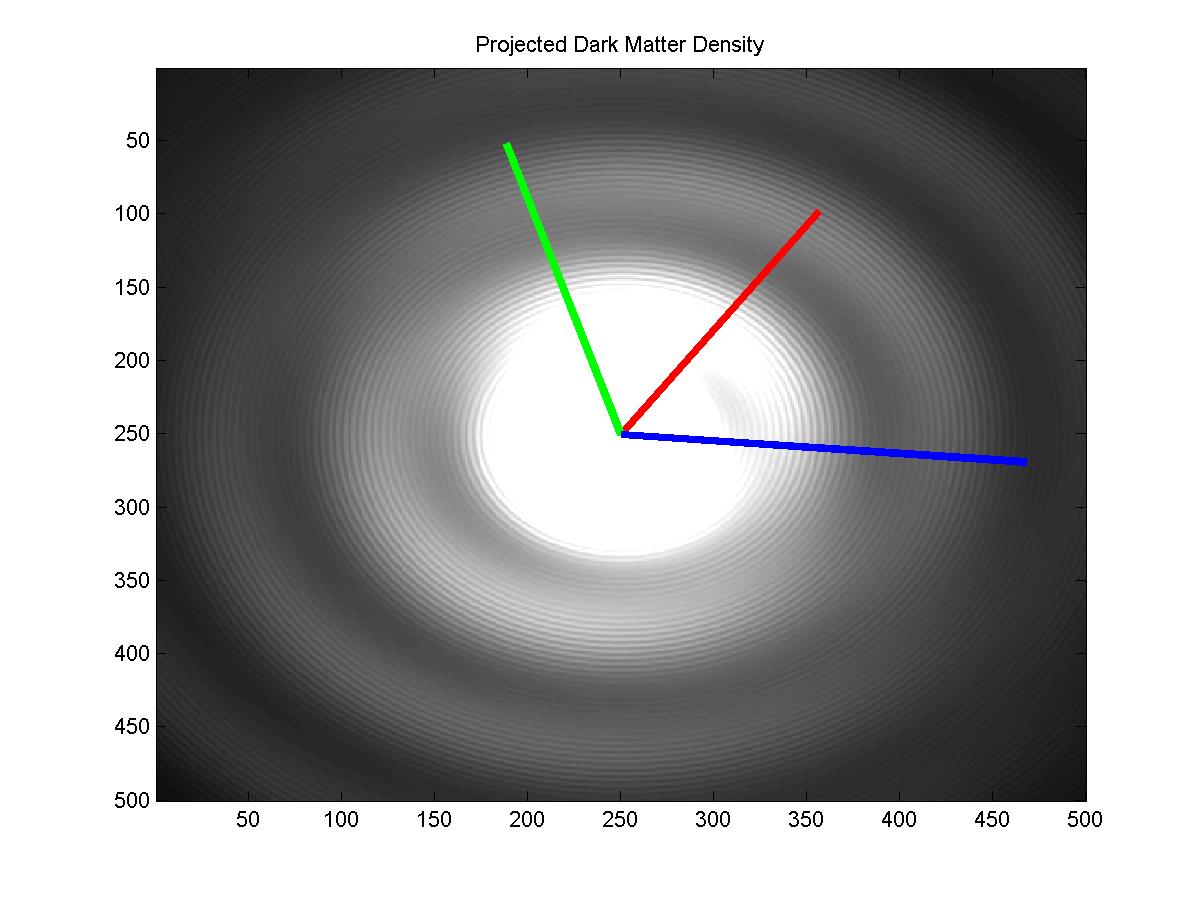}
   \end{center}
   \caption{Projected wave dark matter density, where the $x$-axis is red, the $y$-axis is green, and the $z$-axis is blue.  This is what the wave dark matter would look like if it glowed white, from 6 different point of view.  The $xy$ plane, the $xz$ plane, and the $yz$ plane are shown in the top row.  Images created by running the matlab command shells(500, .1, 100000, 60000, 1, -1, 5, 10), available at the author's ``Wave Dark Matter Web Page'' at http://www.math.duke.edu/\texttildelow bray/darkmatter/darkmatter.html.} \label{P6}
\end{figure}

\section{Wave Dark Matter in Elliptical Galaxies}\label{shells}

Spacetimes associated with galaxies are well approximated by the Minkowski spacetime.  In this paper we will focus on solutions to the Klein-Gordon equation on the Minkowski spacetime.  Solutions in this case are qualitatively similar to solutions to the full Einstein-Klein-Gordon equations in many ways, for example in spherical symmetry.  One main difference is that there are solutions with finite total mass for the Einstein-Klein-Gordon equations which are qualitatively similar to solutions in the Minkowski case, but only up to a certain radius representing the point at which Einstein-Klein-Gordon equation solutions begin to decay exponentially but Klein-Gordon equation solutions on the Minkowski spacetime do not (but instead oscillate forever).  As explained in section 4 of \cite{DMSG}, qualitatively reasonable approximations to solutions to the Einstein-Klein-Gordon equations in equations \ref{eqn:EE} and \ref{eqn:KG} can be found by simply cutting off, outside a large radius, solutions to the Klein-Gordon equation on the Minkowski spacetime.

Using the standard spherical coordinate conventions that
\begin{eqnarray*}
   x &=& r \sin\theta \cos\phi \\
   y &=& r \sin\theta \sin\phi \\
   z &=& r \cos\theta
\end{eqnarray*}
the Klein-Gordon equation on the Minkowski spacetime in spherical coordinates is
\begin{equation}
   \left\{-\frac{\partial^2}{\partial t^2} + \frac{\partial^2}{\partial r^2} + \frac{2}{r} \cdot \frac{\partial}{\partial r}
   + \frac{1}{r^2}\left(
   \frac{\partial^2}{\partial \theta^2} + \cot{\theta} \cdot \frac{\partial}{\partial \theta}
   + \frac{1}{\sin^2{\theta}}\frac{\partial^2}{\partial \phi^2}\right)\right\} f = \Upsilon^2 f.
\end{equation}
Solutions can
then be expanded in terms of spherical harmonics to get solutions
which are linear combinations of solutions of the form
\begin{equation}\label{basis}
   f = A \cos(\omega t) \cdot Y_n(\theta,\phi) \cdot r^n \cdot f_{\omega,n}(r)
\end{equation}
where
\begin{equation}\label{sphericalode}
   f_{\omega,n}''(r) + \frac{2(n+1)}{r} f_{\omega,n}'(r) =
   (\Upsilon^2 - \omega^2) f_{\omega,n}
\end{equation}
and $Y_n(\theta,\phi)$ is an $n$th degree spherical harmonic.  Note
that we require $f_{\omega,n}'(0) = 0$ but have not specified an
overall normalization.  Naturally, to get a complete basis of
solutions we also need to include solutions like the one above but
where $\cos(\omega t)$ is replaced by $\sin(\omega t)$. We will
study real solutions in this paper, but complex solutions are quite
analogous.

The form of the Minkowski spacetime solutions that we will study here is
\begin{equation}\label{angmomsol}
   f = A_0 \cos(\omega_0 t) f_{\omega_0,0}(r) + A_1 \cos(\omega_1 t -
   \phi) \sin{\theta} \cdot r \cdot f_{\omega_1,1}(r)
\end{equation}
where $f_{\omega_0,0}(r)$ and $f_{\omega_1,1}(r)$ satisfy equation
\ref{sphericalode}.  We note these solutions fall into the form of
equation \ref{basis} since both $\cos\phi\sin\theta$ and
$\sin\phi\sin\theta$ are first degree spherical harmonics, namely the Cartesian coordinates
$x$ and $y$ restricted to the unit sphere in $R^3$.
Hence, the above Minkowski spacetime solution is the sum of a
spherically symmetric solution (degree $n=0$) and two degree one
solutions.

The wave dark matter solution represented by equation \ref{angmomsol}
can be seen to be slowly rotating in the $xy$ plane
by making the substitution
\begin{equation}
   \alpha = \phi - \left(\omega_1 - \omega_0\right) t
\end{equation}
into our expression for $f$ in equations \ref{angmomsol} to get
\begin{equation}\label{angmomsol2}
   f = A_0 \cos(\omega_0 t) f_{\omega_0,0}(r) + A_1 \cos(\omega_0 t -
   \alpha) \sin\theta \cdot r \cdot f_{\omega_1,1}(r).
\end{equation}
Note that to the extent that $\alpha$ stays fixed in time, then the
above solution does not rotate and gives a fixed interference
pattern since both terms are oscillating in time with the same
frequency $\omega_0$.  Hence, we see that we get
a dark matter interference pattern which is rotating according to
the formula
\begin{equation}\label{rotationformula}
   \phi_0 = (\omega_1 - \omega_0) t
\end{equation}
with period
\begin{equation}\label{periodformula}
   T_{DM} = \frac{2\pi}{\omega_1 - \omega_0}.
\end{equation}
Note that a different combination of first degree harmonic terms could produce different dynamics, not just rotations.
This raises natural questions about observed shells in ellipticals:  do they ever rotate and, more generally, what are their dynamics?  These are potentially very hard questions because the time scales involved could perhaps be on the order of pattern periods in spiral galaxies, which can be hundreds of millions of years.

Our next goal is to approximate the wave dark matter density $\mu_{WDM}$ due to
this scalar field dark matter solution and then to expand it in
terms of spherical harmonics. From equation \ref{angmomsol2}, we
have that
\begin{eqnarray}
   f &=& \cos(\omega_0 t) \left[A_0 f_{\omega_0,0}(r)
   + A_1 \cos\alpha \sin\theta \cdot r \cdot f_{\omega_1,1}(r)\right] \nonumber \\
   &&+ \sin(\omega_0 t) \left[A_1 \sin\alpha \sin\theta \cdot r \cdot
   f_{\omega_1,1}(r)\right].
   \label{angmomsol3}
\end{eqnarray}

Hence, by equation \ref{eqn:EE}, the wave dark matter density is
\begin{eqnarray}
   \mu_{WDM} &=& \frac{1}{8\pi} G(\partial_t,
   \partial_t) \\
   &\approx& \frac{1}{\Upsilon^2}\left( 2 f_t^2 + (- f_t^2 + |\nabla_x f|^2) \right) +
   f^2 \\
   &\approx& \left( \frac{f_t}{\Upsilon} \right)^2 + f^2
\end{eqnarray}
where we have assumed a long wavelength solution in the
approximation so that spatial derivatives of $f$ are much smaller than the time derivatives.
We will also assume that $\omega_0 \approx \Upsilon \approx
\omega_1$, which allows us to think of
$\alpha$ as being approximately fixed in time. Hence,

\begin{eqnarray}
   \mu_{WDM} &\approx& \left[A_0 f_{\omega_0,0}(r)
   + A_1 \cos\alpha \sin\theta \cdot r \cdot f_{\omega_1,1}(r)\right]^2
   + \left[A_1 \sin\alpha \sin\theta \cdot r \cdot
   f_{\omega_1,1}(r)\right]^2 \label{dmd} \\
   &=& A_0^2 f_{\omega_0,0}(r)^2 + A_1^2 \sin^2\theta \cdot r^2 \cdot
   f_{\omega_1,1}(r)^2 + 2A_0 A_1 \cos\alpha \sin\theta \cdot r \cdot
   f_{\omega_0,0}(r) f_{\omega_1,1}(r) \label{dmd2}
\end{eqnarray}

Note that equation \ref{dmd} only depends on $A_0$, $A_1$, and the two functions $f_{\omega_0,0}(r)$ and $f_{\omega_1,1}(r)$ defined by equation \ref{sphericalode}.  If we define $k_n^2 = \omega_n^2 - \Upsilon^2$, then these two function only depend on $k_0$ and $k_1$, which are roughly the spatial frequencies of $f_{\omega_0,0}(r)$ and $f_{\omega_1,1}(r)$ seen in Figure \ref{inandoutofphase}.  To be physically relevant (more specifically to have group velocities for the wave dark matter much less than the speed of light), $\omega_0$ and $\omega_1$ must be close to $\Upsilon$, resulting in $k_0$ and $k_1$ being small and wavelengths for $f_{\omega_0,0}(r)$ and $f_{\omega_1,1}(r)$ much greater than $1/\Upsilon$.  This observation, combined with precise measurements of the radii of shells in elliptical galaxies thought to be due to wave dark matter, might be able to be used to find lower bounds, though not very sharp, on the value of $\Upsilon$.  This is a good project for another paper.

\subsection{Computer Simulations of the Interleaved Shells in the Dark Matter Density}

The wave dark matter density defined in equation \ref{dmd} has interleaved shells.  The easiest way to visualize these shells is to simply compute examples and look at the pictures generated by Matlab code, as the author has done.  The Matlab algorithm ``shells.m'' is located on the author's ``Wave Dark Matter Web Page'' at http://www.math.duke.edu/\texttildelow bray/darkmatter/darkmatter.html.  All of the computer generated interleaved shells in this paper were created by executing the command line
\vspace{.1in}
\begin{center}
shells(500, .1, 100000, 60000, 1, -1, 5, 10);
\end{center}
In the above command line, $100000$ is the maximum radius to which the dark matter density was computed and $0.1$ was the step size used to solve the o.~d.~e.~in equation \ref{sphericalode}.  $60000$ is the radius of the view displayed in each generated image.  $500$ is the number of pixels in each direction in each image.  $1$ and $-1$ are the values for $A_0$ and $A_1$.  When $|A_0| = |A_1|$, the magnitude of the interleaved shells relative to the overall density is maximized by convention of the overall normalizations of $f_{\omega_0,0}(r)$ and $f_{\omega_1,1}(r)$.
The program then lets the user specify the number of major shells desired, which in this case is $5$ (on each side to a radius of $100000$, or $3$ to a radius of $60000$).  Finally, $10$ is the number of minor shells inside each major shell, as can be seen in Figure \ref{CS1}.  The matlab code uses the requested number of major shells and the requested number of minor shells within each major shell to compute the spatial frequencies $k_0$ and $k_1$ described above.  In the examples in this paper, $k_0 \approx 0.0033$ for a wavelength of about $1900$ and $k_1 \approx 0.0030$ for a wavelength of about $2100$.  One could take these numbers to be in the units of light-years.  Since an overall scaling is allowed, however, the individual values are not relevant; only the relative values are important.

Figure \ref{CS1} shows the wave dark matter density in the $xy$ plane.  The red axis is the $x$-axis, the green axis is the $y$-axis, and in other pictures the blue axis is the $z$- axis.  Note that, ironically, we are coloring the dark matter white in every computer generated image of interleaved shells in the dark matter density.  The next image, Figure \ref{P1}, shows what the dark matter would look like from a distant observer looking down on the $xy$ plane from the point of view of the $z$ axis if the dark matter glowed white everywhere.  These ``projected dark matter densities'' are created by taking the entire three dimensional wave dark matter density and projecting it additively into the chosen plane.
Figures \ref{CS2} and \ref{P2} are the same as Figures \ref{CS1} and \ref{P1} except that the point of view has been changed.  Note that the blue $z$-axis is now visible.  Also note that the image in Figure \ref{P2} is the same as the second image in Figure \ref{F2}, with the first image in Figure \ref{F2} being a dark view of the same wave dark matter density so that more detail in the center of the image may be seen.

All of these computer generated images of interleaved shells in the wave dark matter density should be compared to actual photos of the interleaved shells of elliptical galaxies, such as the image of NGC 474 in Figure \ref{Figure:NGC474}.  Again, there is no way to know at this time if interleaved shells in wave dark matter actually contribute to visible interleaved shells in elliptical galaxies.  This is a question that deserves further study.  Our purpose here is to highlight this intriguing possibility.

Finally, Figure \ref{P6} shows what the wave dark matter density would look like from various angles if the wave dark matter glowed white.  Note that when viewing the wave dark matter density from along the red $x$-axis that the interleaved shells are not visible.  The closer that the point of view is to the $x$-axis, the less pronounced the shells are.  This is important because most elliptical galaxies, like the ones in Figure \ref{F8}, do not have visible shells.  We should also note that we chose $|A_1| = |A_0|$ to maximize the magnitude of the interleaved shells so that they would be easily visible in the computer generated images. Choosing $|A_1| < |A_0|$ makes the interleaved shells as subtle as one wants, until they disappear entirely when $A_1 = 0$.

\subsection{More Analysis}

Now that we have taken advantage of the fact that a ``picture is worth a thousand words'' we return to equations \ref{dmd} and \ref{dmd2} and Figure \ref{inandoutofphase} to see how these interleaved shells may be understood in analytical terms.  For example, we may write
the wave dark matter density from equation \ref{dmd2} as the sum of a spherically symmetric term, an axially symmetric term, and a third term which we will call the interleaved shell density.
\begin{equation}
   \mu_{WDM} = \mu_{Spherical} + \mu_{Axial} + \mu_{Shells}
\end{equation}
where
\begin{eqnarray}
   \mu_{Spherical} &=&  A_0^2 f_{\omega_0,0}(r)^2 \\
       \mu_{Axial} &=& A_1^2 \sin^2\theta \cdot r^2 \cdot f_{\omega_1,1}(r)^2  \\
       \mu_{Shells} &=& 2A_0 A_1 \cos\alpha \sin\theta \cdot r \cdot f_{\omega_0,0}(r) f_{\omega_1,1}(r).
\end{eqnarray}
When $t = 0$, $\alpha = \phi$ so that
\begin{equation}
   \mu_{Shells} = 2A_0 A_1 \cos\phi \sin\theta \cdot r \cdot f_{\omega_0,0}(r) f_{\omega_1,1}(r) = 2A_0 A_1 \cdot x \cdot f_{\omega_0,0}(r) f_{\omega_1,1}(r).
\end{equation}
It follows that the interleaved shell density, which may be positive or negative, is an odd function.  The oddness of $\mu_{Shells}$ explains why the shells are interleaved:  When $\mu_{Shells}$ is a local maximum in the middle of a shell, it is a local minimum at the antipodal point on the other side of the elliptical galaxy.  The shells themselves are created when $f_{\omega_0,0}(r)$ and $f_{\omega_1,1}(r)$ go in and out of phase with one another, as shown in Figure \ref{inandoutofphase}.  When $x$ is positive, a shell is created when the two functions are roughly in phase with one another.  When $x$ is negative, a shell is created when the two functions are roughly out of phase with each other.  Note that the two function are in phase with each other three times in Figure \ref{inandoutofphase}, and that there are three major shells on each side of the origin in the computer generated figures that follow.

\subsection{Conclusion}

Hence we see that wave dark matter might predict interleaved shells in the density of dark matter.  Since interleaved shells are observed in the images of elliptical galaxies, this qualitative similarity is both very intriguing and represents another good reason to continue to study wave dark matter as a serious candidate for dark matter.

This global interleaved shell structure in the wave dark matter density exemplified in Figures \ref{F2}, \ref{CS1}, \ref{P1}, \ref{CS2}, \ref{P2}, and \ref{P6},
while invisible itself, could conceivably play a role in the formation of visible interleaved shells in some
elliptical galaxies, perhaps as exemplified in Figures \ref{F1} and \ref{Figure:NGC474}, through gravity, friction, dynamical friction, and other processes. This conjecture is based primarily on the striking qualitative
similarities in the previously mentioned images, namely that both sets of images have interleaved
shells. The basic idea that regions of increased dark matter density could gravitationally attract more regular visible matter and hence contribute to visible shells is plausible but is also a very subtle question
to study. Resolving this conjecture with a high degree of confidence may require sophisticated
computer simulations based on expert modelings of many different astrophysical processes, making this
a challenging, yet very worthwhile, conjecture to study.

\newpage
\appendix

\section{Comment for the Matlab Function Shells.m}

We include one more comment about the matlab function shells.m which produced all of the computer generated images of the interleaved shells in the wave dark matter density in this paper.  The comment is that the wave dark matter density in shells.m was expressed in terms of spherical harmonics which, while not needed for the discussion in this paper, is needed if one wants to compute the corresponding gravitational potential in the same manner as in \cite{DMSG}.

In order to expand equations \ref{dmd} and \ref{dmd2} into spherical harmonics, we
need to recall that spherical harmonics defined on the unit sphere
are actually restrictions of homogeneous polynomials of the same
degree which are harmonic in $R^3$.
Here is a short
list of homogeneous harmonic polynomials:  degree zero:  $1$, degree
one:  $x,y,z$, degree two:  $x^2 - y^2, 2xy, 3z^2 - r^2, 2xz, 2yz$, where
$r^2 = x^2 + y^2 + z^2$.  It is easy to check that these are all
harmonic in $R^3$ and hence their restrictions to the unit sphere
are spherical harmonics of the same degree.

Since $\cos\alpha\sin\theta = \cos(\phi - \phi_0)\sin\theta$ is a rotated degree one spherical harmonic (the restriction of $x$ to the unit sphere)
and hence a spherical harmonic itself (taking $\phi_0$ to be fixed),
the last term of equation \ref{dmd2} is already in the form
of a spherical harmonic times a function of $r$, as we desire. The
first term is as well, since it is already a function of $r$ and the
zeroth degree spherical harmonic is the constant function one. Hence, it is
only the middle term that we need to put into the desired form. To
do this, we note that
\begin{equation*}
   r^2 \sin^2\theta = r^2 - z^2 = \frac23 r^2 + \frac13 (r^2 - 3z^2),
\end{equation*}
since $z = r \cos\theta$, which successfully expresses this term as the sum of two terms,
each of which is a function of $r$ times a spherical harmonic.
Hence,
\begin{equation}\label{dmdensitysphericalharmonics}
   \mu_{DM} \approx U_0(r) + U_2(r) (3z^2 - r^2) +
   \tilde{U}_1(r) (r \cos\alpha\sin\theta)
\end{equation}
where
\begin{eqnarray}
   U_0(r) &=& A_0^2 f_{\omega_0,0}(r)^2 + \frac23 A_1^2 r^2 f_{\omega_1,1}(r)^2 \nonumber \\
   U_2(r) &=& -\frac13 A_1^2 f_{\omega_1,1}(r)^2 \nonumber \\
   \tilde{U}_1(r) &=& 2 A_0 A_1 f_{\omega_0,0}(r) f_{\omega_1,1}(r).
   \label{sphericalharmoniccomponents}
\end{eqnarray}
These expressions for the wave dark matter potential are the ones used in the matlab function shells.m, available at the author's ``Wave Dark Matter Web Page'' at \\
http://www.math.duke.edu/\texttildelow bray/darkmatter/darkmatter.html.

%-----------------------------------------------------------------------


\newpage
\begin{thebibliography}{99}
\bibliographystyle{plain}


\bibitem{MSBS} A. Bernal, J. Barranco, D. Alic, and C. Palenzuela,
``Multi-state Boson Stars,'' 2009 (http://arxiv.org/abs/0908.2435).

\bibitem{Flat} A. Bernal, T. Matos, D.
N\'{u}\~{n}ez, ``Flat Central Density Profiles from Scalar Field
Dark Matter Halos,'' Report number:  CIEA/03/gr-5
(http://arxiv.org/abs/astro-ph/0303455v3).

\bibitem{BL} G. Bertin and C.C. Lin, ``Spiral Structure in Galaxies: A Density Wave
Theory,'' The MIT Press, 1996.

\bibitem{BHS} G. Bertone, D. Hooper, and J. Silk,
``Particle Dark Matter: Evidence, Candidates and Constraints,''
\emph{Phys. Rept.} 405, 279-390, 2005
(http://arxiv.org/abs/hep-ph/0404175).

\bibitem{BM} J. Binney and M. Merrifield, ``Galactic Astronomy,''
Princeton Series in Astrophysics, Princeton University Press, 1998.

\bibitem{BT} J. Binney and S. Tremaine, ``Galactic Dynamics,''
Princeton Series in Astrophysics, Princeton University Press, 2008.

\bibitem{DMSG} H. Bray, ``On Dark Matter, Spiral Galaxies, and the Axioms of General Relativity,'' AMS Contemporary Mathematics Volume, to appear in 2013, [arXiv:1004.4016], (http://arxiv.org/abs/1004.4016).

\bibitem{BrayParry} H. Bray and A. Parry, ``Modeling Wave Dark Matter in Dwarf Spheroidal Galaxies,'' available on the arXiv in December 2012.

\bibitem{MissingSatellitesProblem} J.S. Bullock, ``Notes on the Missing Satellites Problem,'' XX Canary Islands Winter School of Astrophysics on Local Group Cosmology, Ed. D. Mart\'inez-Delgado, arXiv:1009.4505 [astro-ph.CO], http://arxiv.org/abs/1009.4505.

\bibitem{Cartan} E. Cartan, ``Sur les equations de la graviation
d'Einstein'' J. Math. Pures Appl. I:141-203, 1922.

\bibitem{WIMPSuccess} C. Conroy, R. H. Wechsler, A. V. Kravtsov, ``Modeling Luminosity-Dependent Galaxy Clustering Through Cosmic Time,'' Astrophys. J. 647:201-214, 2006,  	arXiv:astro-ph/0512234, http://arxiv.org/abs/astro-ph/0512234.

\bibitem{CoreCusp} W.J.G. de Blok, ``The Core-Cusp Problem,'' Dwarf Galaxy Cosmology special issue of Advances in Astrophysics, arXiv:0910.3538 [astro-ph.CO], http://arxiv.org/abs/0910.3538.

\bibitem{GM} F. S. Guzman, T. Matos, and  H. Villegas-Brena, ``Scalar dark matter in spiral
galaxies,'' \emph{Rev. Mex. Astron. Astrofis.} 37 (2001) 63-72
(http://arxiv.org/abs/astro-ph/9811143).

\bibitem{WMAP} G. Hinshaw, J. L. Weiland, R. S. Hill, N. Odegard, D. Larson, C. L. Bennett,
J. Dunkley, B. Gold, M. R. Greason, N. Jarosik, E. Komatsu, M. R.
Nolta, L. Page, D. N. Spergel, E. Wollack, M. Halpern, A. Kogut, M.
Limon, S. S. Meyer, G. S. Tucker, E. L. Wright, ``Five-Year
Wilkinson Microwave Anisotropy Probe (WMAP) Observations: Data
Processing, Sky Maps, and Basic Results,'' Astrophys. J. Suppl.
180:225-245, 2009, arXiv:0803.0732v2 [astro-ph]
(http://arxiv.org/abs/0803.0732v2).

\bibitem{Hooper} D. Hooper and E. Baltz, ``Strategies for
Determining the Nature of Dark Matter,'' \emph{Annu. Rev. Nucl.
Part. Sci.} 58, 293-314, 2008 (http://arxiv.org/abs/0802.0702).

\bibitem{JS} S. U. Ji and S.-J. Sin, Phys. Rev. D 50, 3655
(1994).

\bibitem{LaiChoptuik} C. W. Lai, M. W. Choptuik, ``Final Fate of Subcritical Evolutions of Boson Stars,'' arXiv:0709.0324 (http://arxiv.org/abs/0709.0324).

\bibitem{JWLee} J.-W. Lee, ``Is Dark Matter a BEC or Scalar Field?''
\emph{Journal of the Korean Physical Society}, Vol. 54, No. 6, 2622,
June 2009 (http://arxiv.org/abs/0801.1442).

\bibitem{LeeKoh1992} J.-W. Lee and I.-G. Koh, ``Galactic Halo as a
Soliton Star,'' Abstracts, bulletin of the Korean Physical Society,
10 (2) (1992).

\bibitem{LeeKoh1996} J.-W. Lee and I.-G. Koh, Phys. Rev. D 53, 2236
(1996).

\bibitem{LiChen} N. Li, D.-M. Chen, ``Cusp-core Problem and Strong Gravitational Lensing,''  	 Res. Astron. Astrophys. 9:1173-1184 (2009), arXiv:0905.3041 [astro-ph.CO], http://arxiv.org/abs/0905.3041.

\bibitem{Lovelock} D. Lovelock, ``The four-dimensionality of space and the Einstein
tensor,'' \emph{J. Mathematical Phys.} 13:874-876, 1972.

\bibitem{Matos1} T. Matos, L. A. Urena-Lopez, 2000, Class. quantum
Grav. 17, L75, arXiv:astro-ph/0004332, http://arxiv.org/abs/astro-ph/0004332.

\bibitem{Matos2} T. Matos, L. A. Urena-Lopez, 2001, Phys. Rev. D
63, 063506.

\bibitem{McGaugh2011} S. McGaugh, ``The Baryonic Tully-Fisher Relation of Gas Rich Galaxies as a Test of LCDM and MOND,'' Astronomical Journal, July 14, 2011, arXiv:1107.2934 [astro-ph.CO] (http://arxiv.org/abs/1107.2934).

\bibitem{McGaugh2009} S. McGaugh, J. M. Schombert, W.J.G. de Blok, M. J. Zagursky, ``The Baryon Content of Cosmic Structures,'' ApJ Letters, November 13, 2009, arXiv:0911.2700 [astro-ph.CO] (http://arxiv.org/abs/0911.2700).

\bibitem{NFW1} J.F. Navarro, C.S. Frenk, S.D.M. White, 1996, ApJ, 462, 563.

\bibitem{NFW2} J.F. Navarro, C.S. Frenk, S.D.M. White, ``A Universal Density Profile from Hierarchical Clustering,'' 1997, ApJ, 490: 493-508, arXiv:astro-ph/9611107, http://arxiv.org/abs/astro-ph/9611107.

\bibitem{ONeill} B. O'Neill, ``Semi-Riemannian Geometry with
Applications to Relativity,'' Academic Press, 1983.

\bibitem{Ostriker} J. Ostriker, ``Astronomical Test of the Cold Dark
Matter Scenario,'' \emph{Annual Review of Astronomy and
Astrophysics}, Volume 31, 689-716, 1993
(http://adsabs.harvard.edu/abs/1993ARA\%26A..31..689O).

\bibitem{Parry1} A. R. Parry, ``A Survey of Spherically Symmetric Spacetimes'' (2012), available at http://arxiv.org/abs/1210.5269.

\bibitem{Parry2} A. R. Parry, ``Spherically Symmetric Static States of Wave Dark Matter,'' (2012), available on the arXiv in December 2012.

\bibitem{AU2} S. Perlmutter et al. (The Supernova Cosmology Project) (1999) ``Measurements of Omega and Lambda from 42 high redshift supernovae,'' Astrophysical J. 517 (2): 565–86. arXiv:astro-ph/9812133. Bibcode 1999ApJ...517..565P. doi:10.1086/307221 (http://arxiv.org/abs/astro-ph/9812133).

\bibitem{AU1} Adam G. Riess et al. (Supernova Search Team) (1998) ``Observational evidence from supernovae for an accelerating universe and a cosmological constant,'' Astronomical J. 116 (3): 1009–38. arXiv:astro-ph/9805201. Bibcode 1998AJ....116.1009R. doi:10.1086/300499 (http://arxiv.org/abs/astro-ph/9805201).

\bibitem{DMAW} P. Salucci, C. F. Martins, A. Lapi, ``DMAW 2010 LEGACY the Presentation Review: Dark Matter in Galaxies with its Explanatory Notes,'' presentation at http://www.sissa.it/ap/dmg/dmaw\_presentation.html, notes at arXiv:1102.1184 [astro-ph.CO]
    (http://arxiv.org/abs/1102.1184).

\bibitem{SM} F. Schunck and E. Mielke, ``Topical Review:  General
Relativistic Boson Stars,'' \emph{Class. Quant. Grav.} 20,
R301-R356, 2003 (http://arxiv.org/abs/0801.0307).

\bibitem{SKM} R. Sharma, S. Karmakar, and S. Mukherjee, ``Boson star and dark
matter,'' Dec. 2008 (http://arxiv.org/abs/0812.3470).

\bibitem{Sin1992} S.-J. Sin, Phys. Rev. D 50, 3650 (1994).

\bibitem{Trimble} V. Trimble, ``Existence and Nature of Dark Matter
in the Universe,'' \emph{Annual Review of Astronomy and
Astrophysics}, Volume 25, 425-472, 1987
(http://adsabs.harvard.edu/abs/1987ARA\%26A..25..425T).

\bibitem{Vermeil} H. Vermeil, ``Notiz uber das mittlere Krummungsmass einer
n-fach ausgedehnten Riemann'schen Mannigfaltigkeit,'' 1917.

\bibitem{Weyl} Weyl, ``Space, time, matter,'' Dover, 1922.

\bibitem{Gilmore} R. F. G. Wyse, G. Gilmore, ``The Observed Properties of Dark Matter on Small Spatial Scales,'' Proceedings of IAU Symposium 244, ``Dark Galaxies and Lost Baryons'', eds J. Davies and M. Disney, arXiv:0708.1492 [astro-ph], http://arxiv.org/abs/0708.1492.

\end{thebibliography}
\end{document}